\documentclass[%
reprint, 
superscriptaddress, 
showpacs,
amsmath, amssymb,
aps,
prl
]{revtex4-2}
\usepackage{graphicx}
\usepackage{dcolumn}
\usepackage{bm}
\usepackage{upgreek}
\graphicspath{{figure/}{figsgaoerb/}}
\usepackage{hyperref}
\usepackage{braket}
\usepackage{booktabs, array, mathptmx, float, tabularx, booktabs, lipsum, amsmath,multirow}
\hypersetup{colorlinks=true, citecolor=blue, urlcolor=blue, linkcolor=blue}
\usepackage{color}
\begin{document}
\title{New Constraints on Exotic Spin-Spin-Velocity-Dependent Interactions with Solid-State Quantum Sensors} 

\author{Yue Huang}
\affiliation{CAS Key Laboratory of Microscale Magnetic Resonance and School of Physical Sciences, University of Science and Technology of China, Hefei 230026, China}  
\affiliation{CAS Center for Excellence in Quantum Information and Quantum Physics, University of Science and Technology of China, Hefei 230026, China}  

\author{Hang Liang}
\affiliation{CAS Key Laboratory of Microscale Magnetic Resonance and School of Physical Sciences, University of Science and Technology of China, Hefei 230026, China} 
\affiliation{CAS Center for Excellence in Quantum Information and Quantum Physics, University of Science and Technology of China, Hefei 230026, China}

\author{Man Jiao}
\email{jm2012@ustc.edu.cn}
\affiliation{CAS Key Laboratory of Microscale Magnetic Resonance and School of Physical Sciences, University of Science and Technology of China, Hefei 230026, China}
\affiliation{CAS Center for Excellence in Quantum Information and Quantum Physics, University of Science and Technology of China, Hefei 230026, China}

\author{Pei Yu}
\affiliation{CAS Key Laboratory of Microscale Magnetic Resonance and School of Physical Sciences, University of Science and Technology of China, Hefei 230026, China} 
\affiliation{CAS Center for Excellence in Quantum Information and Quantum Physics, University of Science and Technology of China, Hefei 230026, China}

\author{Xiangyu Ye}
\affiliation{CAS Key Laboratory of Microscale Magnetic Resonance and School of Physical Sciences, University of Science and Technology of China, Hefei 230026, China}
\affiliation{CAS Center for Excellence in Quantum Information and Quantum Physics, University of Science and Technology of China, Hefei 230026, China}

\author{Yijin Xie}
\affiliation{Institute of Quantum Sensing and School of Physics, Zhejiang University, Hangzhou 310027, China}

\author{Yi-Fu Cai}
\affiliation{CAS Key Laboratory for Researches in Galaxies and Cosmology, School of Astronomy and Space Science, University of Science and Technology of China, Hefei, Anhui 230026, China}
\affiliation{Department of Astronomy, School of Physical Sciences, University of Science and Technology of China, Hefei, Anhui 230026, China}

\author{ Chang-Kui Duan}
\affiliation{CAS Key Laboratory of Microscale Magnetic Resonance and School of Physical Sciences, University of Science and Technology of China, Hefei 230026, China}
\affiliation{CAS Center for Excellence in Quantum Information and Quantum Physics, University of Science and Technology of China, Hefei 230026, China}
\affiliation{Hefei National Laboratory, University of Science and Technology of China, Hefei 230088, China}

\author{Ya Wang}
\affiliation{CAS Key Laboratory of Microscale Magnetic Resonance and School of Physical Sciences, University of Science and Technology of China, Hefei 230026, China}
\affiliation{CAS Center for Excellence in Quantum Information and Quantum Physics, University of Science and Technology of China, Hefei 230026, China}
\affiliation{Hefei National Laboratory, University of Science and Technology of China, Hefei 230088, China}

\author{Xing Rong}
\email{xrong@ustc.edu.cn}
\affiliation{CAS Key Laboratory of Microscale Magnetic Resonance and School of Physical Sciences, University of Science and Technology of China, Hefei 230026, China}
\affiliation{CAS Center for Excellence in Quantum Information and Quantum Physics, University of Science and Technology of China, Hefei 230026, China}
\affiliation{Hefei National Laboratory, University of Science and Technology of China, Hefei 230088, China}

\author{Jiangfeng Du}
\email{djf@ustc.edu.cn}
\affiliation{CAS Key Laboratory of Microscale Magnetic Resonance and School of Physical Sciences, University of Science and Technology of China, Hefei 230026, China}
\affiliation{CAS Center for Excellence in Quantum Information and Quantum Physics, University of Science and Technology of China, Hefei 230026, China}
\affiliation{Institute of Quantum Sensing and School of Physics, Zhejiang University, Hangzhou 310027, China}
\affiliation{Hefei National Laboratory, University of Science and Technology of China, Hefei 230088, China}

\date{\today}

\begin{abstract}
	We report new experimental results on exotic spin-spin-velocity-dependent interactions between electron spins.
	We designed an elaborate setup that is equipped with two nitrogen-vacancy (NV) ensembles in diamonds. One of the NV ensembles serves as the spin source, while the other functions as the spin sensor. By coherently manipulating the quantum states of two NV ensembles and their relative velocity at the micrometer scale, we are able to scrutinize exotic spin-spin-velocity-dependent interactions at short force ranges. For a T-violating interaction, $V_6$, new limits on the corresponding coupling coefficient, $f_6$, have been established for the force range shorter than 1 cm. For a P,T-violating interaction, $V_{14}$, new constraints on the corresponding coupling coefficient, $f_{14}$, have been obtained for the force range shorter than 1 km.
\end{abstract}
\maketitle

\begin{figure}[htbp!]
	\centering
	\includegraphics[width=1\columnwidth]{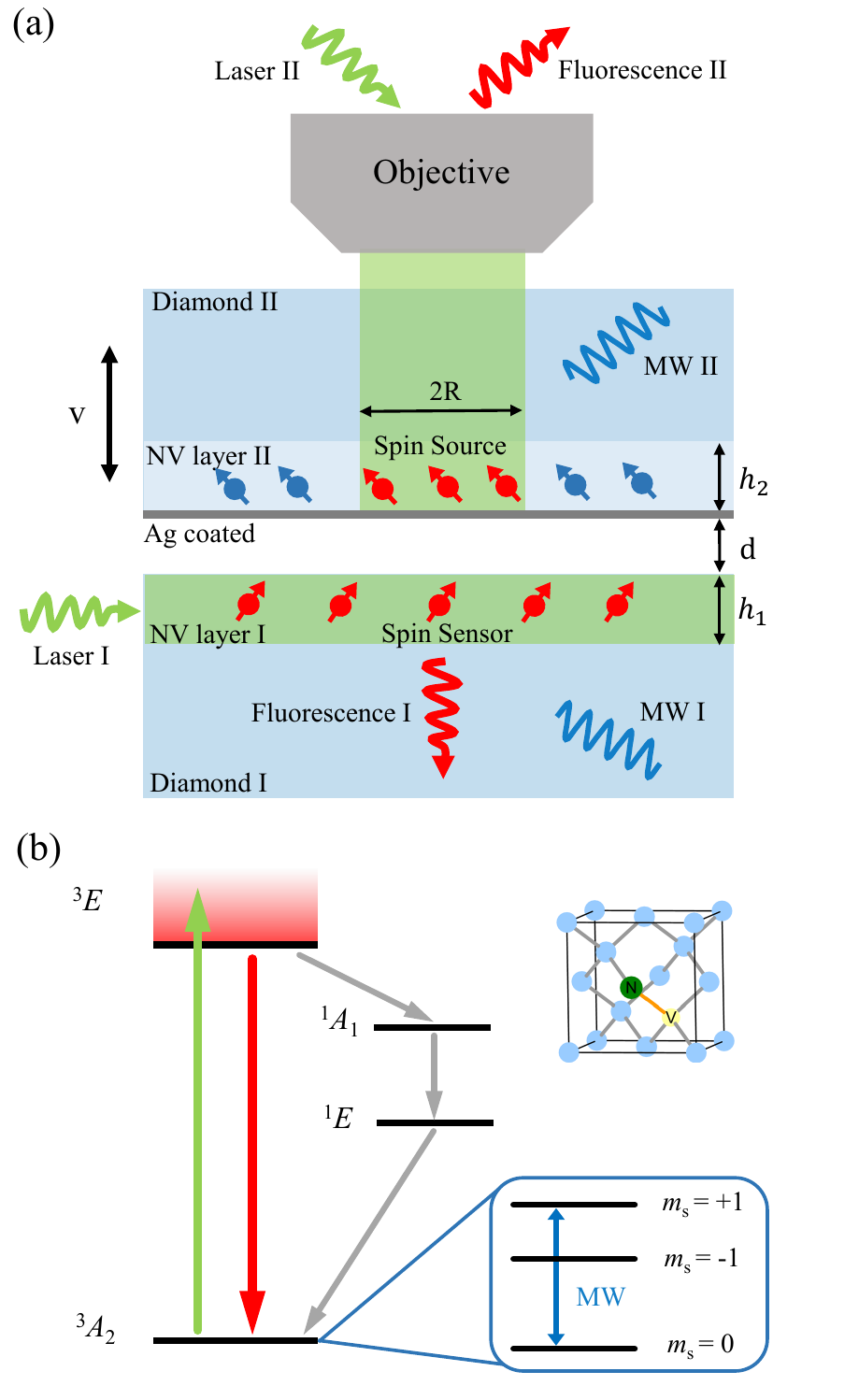}
	\caption{(a) Schematic experimental setup. The spin sensor consists of a thin layer of NV centers at the surface of a diamond chip (labeled as Diamond I). The spin source is another thin NV layer on a diamond chip (labeled as Diamond II). The laser and microwave applied on the spin sensor (source) are labeled as Laser I (II) and MW I (II), respectively. Laser I and Laser II illuminated the spin sensor and spin source from the side and top, respectively. There was a silver layer isolating the two optically detected systems. The spin source was modulated to vibrate with velocity $\vec{v}$ perpendicular to the diamond surface.  (b) Energy-level diagram and atomic structure of NV center in diamond. NV electronic spin states were excited and initialized by green laser and detected via spin-state-dependent red fluorescence. The triplet ground-state spin transitions can be excited by resonant microwaves.}
	\label{SAsetup}
\end{figure}

Ultralight new bosons ($m_b\lesssim$ 1 eV/$c^2$)\cite{jackson-kimball-probing-2023} beyond the Standard Model are proposed to explain mysteries of modern physics, such as strong CP problem\cite{peccei1977constraints,weinberg1978new,wilczek1978problem}, the hierarchy problem\cite{Graham2015b}, and the composition of dark matter\cite{Kim2010}.
It's predicted that these hypothetical bosons, including axions\cite{PeterSvrcek_2006}, familons\cite{PhysRevLett_49_1549}, paraphotons\cite{Dobrescu2005}, Z$'$ bosons\cite{okada_dark_2020}, etc., can serve as mediators of exotic interactions between fermions\cite{Moody1984,dobrescu2006spin}.
Such spin-0 or spin-1 boson exchanges within a Lorentz-invariant quantum field theory can be categorized into fifteen exotic spin-dependent interactions, which enable methodical exploration with astrophysical and laboratory searches \cite{dobrescu2006spin,Fadeev2019a}.

With recent advances in precision measurement, spin based sensors play a vital role in table-top experiments searching for the exotic spin-dependent interactions \cite{safronova_search_2018}, such as the ion trap\cite{Street2015}, atomic magnetometer\cite{Ji2018RPL,Su2021,Wang2022}, scanning probe microscope\cite{Ding2020}, NV centers in diamond\cite{Rong2018,Jiao2021c,liang_new_2022,wu_improved_2023} and polarized torsion pendulum\cite{Heckel2013,Terrano2015}. 
Exotic Spin-Spin-Velocity-Dependent Interactions (SSVDIs) can be mediated by spin-1 bosons including the new massless paraphotons and light Z$'$ bosons\cite{dobrescu2006spin}. In contrast to interactions introduced by spin-0 bosons, interactions mediated by new spin-1 bosons can avoid astrophysical constraints due to potential loopholes in astrophysical and cosmological limits\cite{dobrescu2006spin,jain2006evading}, making direct laboratory searching important and necessary. While static exotic spin-spin interactions have been strictly constrained over a broad range of distance scales\cite{Heckel2013,Terrano2015,Rong2018,Almasi2020,Ficek2017A,Hunter2013}, the investigation of SSVDIs is less extensive, especially in the force range below a centimeter\cite{Ji2018RPL,Hunter2014}. The experimental search at short force ranges remains unexplored due to the challenges in coherently steering quantum states of electron spins, high precision magnetic sensing and spatial position modulation at micrometer scale.   

In this work, we experimentally investigated SSVDIs between polarized electrons utilizing two individual ensemble-NV-diamonds. One type of the SSVDIs whose potential following the notation in Ref.\cite{dobrescu2006spin} can be given as
\begin{equation}
	\begin{aligned}
		V_{6}=&-f_{6}\frac{{\hbar}^2}{4\pi m_e c}[(\hat{\sigma_1}\cdot\vec{v})(\hat{\sigma_2}\cdot\hat{r})~\\&+~(\hat{\sigma_1}\cdot\hat{r})(\hat{\sigma_2}\cdot\vec{v})] \left(\frac{1}{\lambda r}+\frac{1}{r^2}\right)e^{-r/\lambda},
	\end{aligned}
	\label{V6}
\end{equation}	
where $f_6$ is the dimensionless coupling coefficient, $\hat{\sigma_1}$ and $\hat{\sigma_2}$ are the unit spin vectors of the two interacting fermions, respectively. $\boldsymbol{v}$ is the relative velocity between them, $r=|\boldsymbol{\vec{r}}|$ is the displacement and $\boldsymbol{\hat{r}}$ is the unit displacement vector. $\hbar$ is the reduced Planck’s constant, $c$ is the speed of light in vacuum and $ m_e$ is the mass of the electron. $\lambda=\hbar/ m_b c$ is the interaction range determined by the mass of the mediated new boson $ m_b$. The SSVDI can be induced by the exchange of virtual Z$'$ bosons\cite{dobrescu2006spin}, which are motivated by various theoretical scenarios of beyond-the-Standard-Model physics, and are candidates for dark matter\cite{langacker-physics-2009}. Laboratory searching for the SSVDI offers a promising avenue to further our understanding of Z$'$ bosons fundamental physics.  
The exotic interaction can be characterized as an effective magnetic field acting on NV electron spins. The effective magnetic field generated from the spin source on the spin sensor is:

\begin{equation}
	\begin{aligned}
		B_{6}=&-f_{6}\frac{\hbar}{2\pi m_e \gamma c}[(\hat{\sigma_1}\cdot\vec{v})(\hat{\sigma_2}\cdot\hat{r})\\&+(\hat{\sigma_1}\cdot\hat{r})(\hat{\sigma_2}\cdot\vec{v})] \left(\frac{1}{\lambda r}+\frac{1}{r^2}\right)e^{-r/\lambda},
	\end{aligned}
	\label{B6}
\end{equation}	
where $\gamma$ is the gyromagnetic ratio of the electron spin.
In this work, we utilized  an NV ensemble as a magnetometer to search for the possible magnetic field due to the SSVDI from another NV ensemble, which acts as a polarized electron spin source. 
Compared with previously used electron spin sources such as SmCo$_5$ magnets\cite{Ji2018RPL}, the electron spin states of NV ensembles can be modulated instantaneously and efficiently via optical and microwave pulses. Moreover, the small geometry size of the ensemble-NV-diamonds enables close proximity between the spin sensor and spin source, which is essential for detecting exotic spin-dependent interactions at short force ranges.
\begin{figure*}[htbp!]
	\centering
	\includegraphics[width=2\columnwidth]{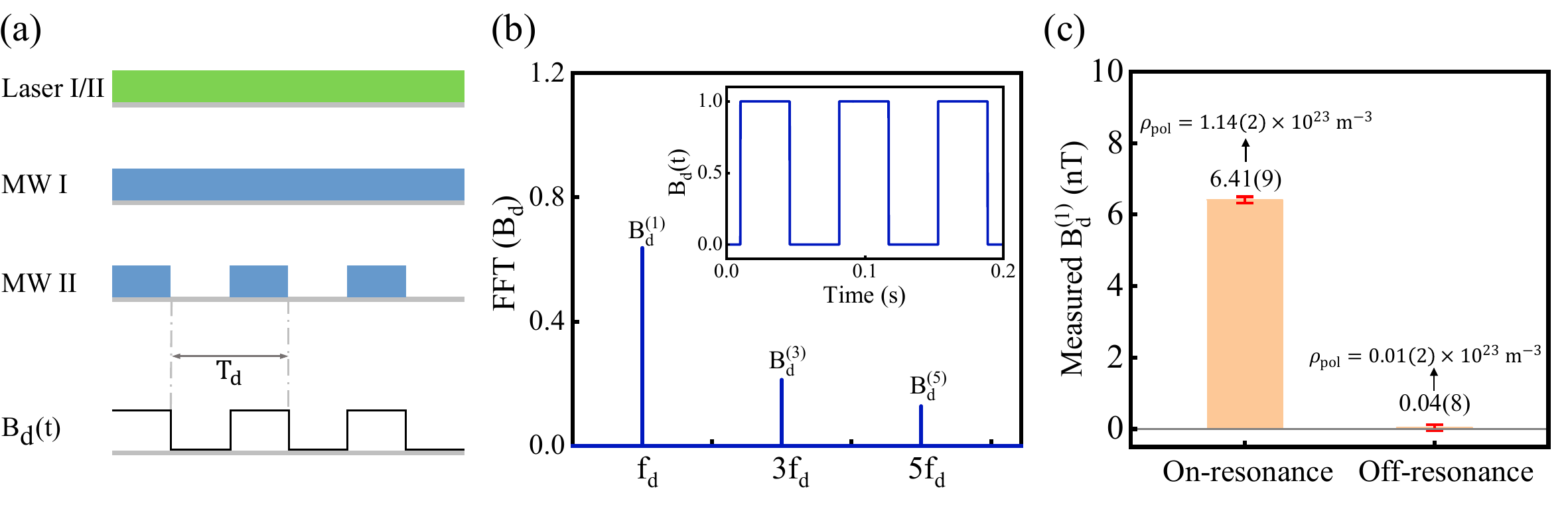}
	\caption{Measurement of the polarized spin density of the spin source. (a) The laser and the microwave sequences applied on the spin sensor and the spin source as well as the time evolution of the magnetic dipole field $B_{\rm d}$. $T_{\text{d}} = 1/f_{\text{d}}$. (b) Fourier transformation spectrum of $B_\text{d}$. Inset: square-wave signal of $B_\text{d}$ in time domain. (c) Experimental results of the measured magnetic dipole fields and the measured polarized spin densities under on-resonance and off-resonance conditions.} 
	\label{Dipole}
\end{figure*}

The geometric schematic of our setup is shown in Fig. \ref{SAsetup}(a). We used two $660\times661\times574~\upmu$m$^3$ diamond crystals (labeled with Diamond I and II in Fig. \ref{SAsetup}) with $\langle$100$\rangle$ oriented surfaces. NV centers with a concentration being 14 (22) ppm were doped within a thin layer $h_1$ ($h_2$) = 23 $\upmu$m at the Diamond I (II) surface. The high concentration NV centers in the two diamonds were utilized as the spin sensor (Diamond I) and the spin source (Diamond II) for SSVDIs, respectively. The spin source was carefully placed above the spin sensor at the position where the distance between the two diamonds was $d$ being 18.5 $\upmu$m. To separately manipulate and read out the spin states of the spin sensor and the spin source, we constructed two sets of laser and fluorescence collection channels. The spin sensor was illuminated by a 532-nm laser with a beam diameter of about 40 $\upmu$m via the flank of the diamond. The red fluorescence emitted from the spin sensor was collected via a compound parabolic concentrator below diamond I. The other green laser was sent through an objective above to excite the spin source with a spot radius $R$ being 52 $\upmu$m. The fluorescence from the spin source was collected by the same objective. A 500-nm-silver layer was fabricated on Diamond II to isolate the two laser beams as well as fluorescence from two NV layers. The sensing area of the spin sensor as well as the polarized region of the spin source were monitored by an upper camera according to fluorescence images (see Appendix \ref{SM1} for details). To investigate SSVDIs, the spin source was modulated by a piezoelectric bender to vibrate at $\boldsymbol{v}$ with frequency $f_{\text{vib}}$ = 1.337 kHz.

As shown in Fig. \ref{SAsetup}(b), NV ground state is an electron spin triplet state with three spin states $|m_s=0\rangle$ and $|m_s=\pm 1\rangle$\cite{doherty_NV_2013}. A static magnetic field $B_0$ being 94 gauss was applied along the NV symmetry axis of the spin sensor to separate $|m_s=\pm 1\rangle$ spin states. We positioned the two diamonds with a relative angle of 54$^{\circ}$ rotated along the vertical direction, such that the projection of the bias magnetic field $B_0$ onto the NV axis differed for each sample. It resulted in the difference between the $|m_s=0\rangle\to|m_s=+1\rangle$ transition frequencies of the two NV ensembles, which enabled us to control them independently (see Appendix \ref{SM1} for details). Microwaves of two distinct frequencies (labeled as MW I and MW II in Fig. \ref{SAsetup}) were delivered via a double-split-ring-resonator to manipulate the spin states of the spin sensor and the spin source, respectively.

The spin sensor is an ensemble-NV-diamond magnetometer in this work. A typical continuous-wave method was carried out with the $|m_s=0\rangle \to |m_s=+1\rangle$ transition of the sensor, wherein laser and microwave field were continuously applied\cite{XIE2021,BarryPNA2016}. 
We applied frequency modulation to the microwave on the spin sensor, encoding the magnetic-field information in a band around the modulation frequency. The laser fluctuation was also recorded for noise cancellation. The magnetic sensitivity of the ensemble-NV-diamond magnetometer is 2 nT/$\sqrt{\text{Hz}}$ within the frequency range from 1 to 2 kHz (see Appendix \ref{SM1} for details).

Prior to searching for the exotic interactions, the polarized spin density of the spin source was obtained by  measuring the magnetic dipole-dipole interaction between the spin sensor and the spin source. Figure. \ref{Dipole}(a) shows schematically the laser and the microwave sequences applied on the spin sensor and spin source. The magnetic polarization of the spin source was modulated by periodically switching MW II via a microwave switch with 50$\%$ duty cycle at frequency being 14 Hz. When the laser continuously pumped the spin source to $|m_s=0\rangle$, nonzero magnetic polarization of the electron spin can be realized with the resonant microwave being turned on.  Since the polarization changes fast during the switching process of MW II, the magnetic field $B_{\text{d}}$ sensed by the spin sensor due to the magnetic dipole-dipole interaction can be characterized as a square wave (see Appendix \ref{SM2} for details). Thus it can be decomposed into a series of odd sinusoidal harmonics:
\begin{equation}
	\begin{aligned}
		B_{\text{d}}&=-\frac{\mu_0 \gamma \hbar}{8\pi}\frac{1}{V_\text{I}}\rho_{\text{pol}}\int_{V_\text{I}}dV\int_{V_{\text{II}}}dV'\frac{3(\hat{\sigma_1}\cdot\hat{r})(\hat{\sigma_2}\cdot\hat{r})-\hat{\sigma_1}\cdot\hat{\sigma_2}}{r^3}  \\
		&=\sum_{n=\rm odd}^{\infty} B^{(n)}_{\text{d}} \sin(2\pi nf_{\text{d}}t),
	\end{aligned}
	\label{Bdipole}
\end{equation}
where $V_\text{I}$ ($V_{\text{II}}$) stands for the integration volume of the spin sensor (spin source), $\mu_0$ is the vacuum permeability, $\rho_{\text{pol}}$ is the polarized spin density of the spin source and $B^{(n)}_{\text{d}}$ is the $n$-th Fourier coefficient of $B_\text{d}$. The frequency spectrum of $B_\text{d}$ is shown in Fig. \ref{Dipole}(b), including components at $f_{\text{d}}$, 3$f_{\text{d}}$, 5$f_{\text{d}}$, etc. We extracted the amplitude of the first-order harmonic $B^{(1)}_{\text{d}}$ with a lock-in amplifier at $f_{\text{d}}$. As shown in Fig. \ref{Dipole}(c), when MW II was on-resonance, $B^{(1)}_{\text{d}}$ was measured to be 6.41(9) nT. When MW II was off-resonance, the result presented a zero signal. The polarized spin source density $\rho_{pol}$ was then obtained to be ($1.14\pm 0.02)\times10^{23}$ m$^{-3}$ when MW II was on-resonance.

\begin{figure}[htbp!]
	\centering
	\includegraphics[width=1\columnwidth]{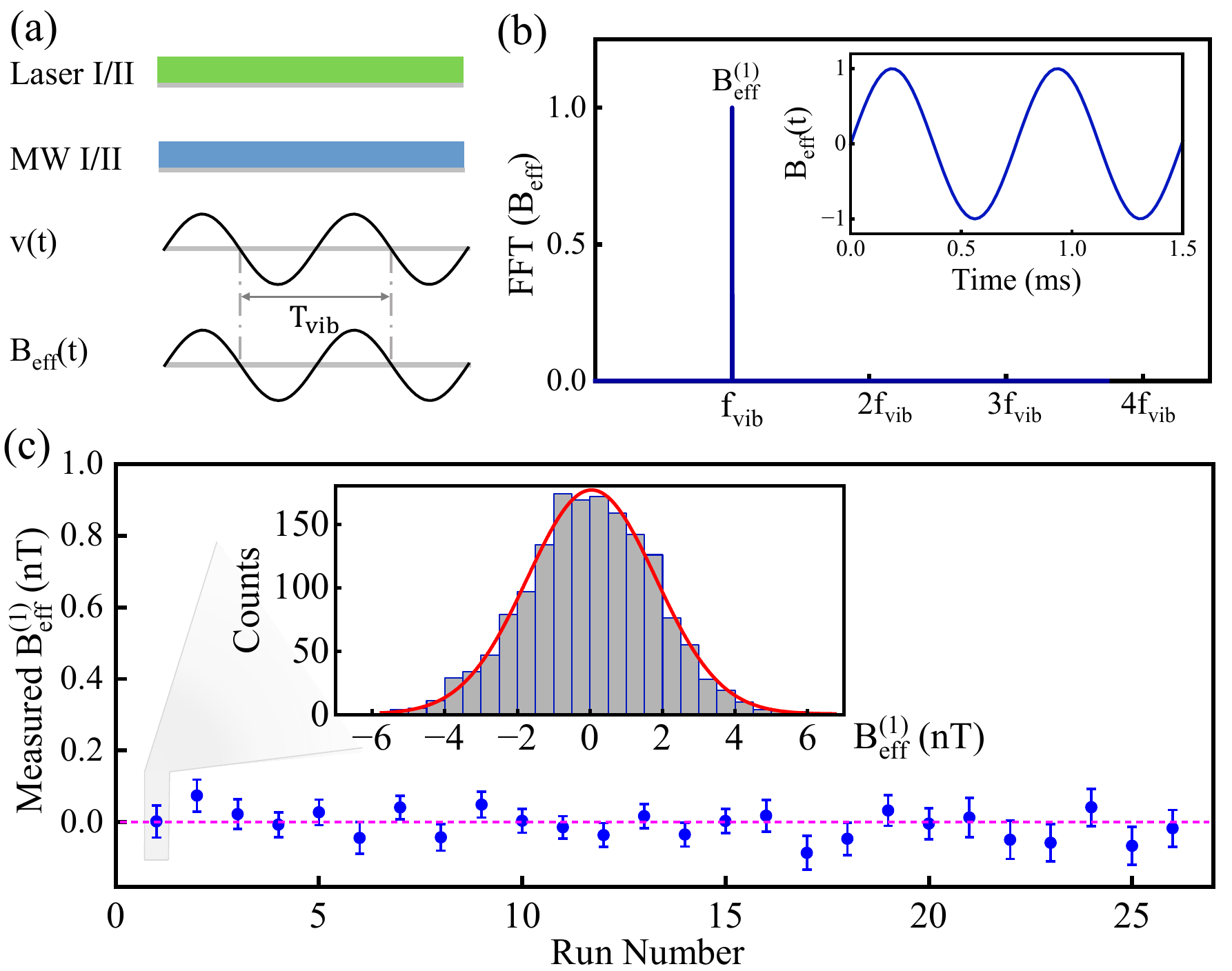}
	\caption{Experimental searching for the SSVDIs. (a) The experimental sequences applied on the spin sensor and spin source, as well as the time evolution of velocity $v$ and the possible effective magnetic field $B_{\text{eff}}$. $T_{\text{vib}}=1/f_{\text{vib}}$. (b) Fourier transformation spectrum of $B_{\text{eff}}$. Inset: calculated $B_{\text{eff}}$ in time domain. (c) Experimental results of the measured effective magnetic field induced by SSVDIs. Each point and its error bar represent the average and the standard error of one-hour dataset. The dashed magenta line marks the zero value of $B^{(1)}_{\text{eff}}$. The top inset shows the histogram of experimental results for the first one-hour dataset, in which the red solid line indicates a valid fit to the Gaussian distribution.} 
	\label{SAresult}
\end{figure}

The experimental sequences to detect the SSVDIs are shown in Fig. \ref{SAresult}(a), together with the time evolution of velocity $v$ and corresponding effective magnetic field $B_{\text{\rm eff}}$. Continuous application of MW II maintained the polarization of the spin source in a steady status, which enabled long-term stable searching. Since the spin source vibrates at a fixed frequency $f_{\rm vib}$=1.337 kHz, the velocity of the spin source can be expressed as $v(t)=2\pi f_{\rm vib} A \sin(2\pi nf_{\rm vib}t)$, where $A$ = 36.7 nm is the vibration amplitude measured by a commercial Doppler interferometer. The effective magnetic field sensed by the ensemble-NV-diamond magnetometer due to $V_6$ is:  
\begin{equation}
	\begin{aligned}
		B_{\text{\rm eff}}&=\frac{1}{V_\text{I}}\rho_{\text{pol}}\int_{V_\text{I}}dV\int_{V_{\text{II}}}dV'B_{6} \\ 
		&=\sum_{n=1}^{\infty} B^{(n)}_{\text{eff}} \sin(2\pi nf_{\text{vib}}t),
	\end{aligned}
	\label{Beff}
\end{equation}
where $B^{(n)}_{\text{eff}}$ is the $n$-th Fourier coefficient of $B_{\text{eff}}$. Based on numerical simulations, the field strength primarily lies in the first-order harmonic component, as shown in Fig. \ref{SAresult}(b) (see Appendix \ref{SM3} for details). The amplitude of the first-order harmonic $B^{(1)}_{\text{eff}}$ was extracted by a lock-in amplifier with demodulation frequency being $f_{\text{vib}}$. After calibration of the phase $\phi$ of the demodulation reference signal, the velocity-dependent signal $B^{(1)}_{\text{eff}}$ and displacement-dependent signal correspond to the quadrature channel and in-phase channel of the lock-in amplifier, respectively (see Appendix \ref{SM3} for details). To eliminate the effects of magnetic dipole-dipole interactions, previous experiments usually employed magnetic field shielding \cite{Ji2018RPL}. This resulted in a considerable distance between the spin source and the spin sensor. In our experiment, the magnetic field shielding is not required due to the following reason. Though the magnetic dipole-dipole interaction between the vibrating spin source and the sensor produces a magnetic field on the sensor, our setup responses only to the magnetic field with a center frequency of $f_{\text{vib}}$ and the DC component of the field does not contribute to the signal collected by our setup. The strength of the AC component of the field due to the magnetic dipole-dipole interaction is negligble and in phase with displacement, which is orthogonal to the signal due to the SSVIDs. Therefore, we can search for the possible signal due to SSVIDs without the influence of the magnetic dipole-dipole interactions and can get rid of the magnetic shielding.

\begin{figure*}[htbp!]
	\centering
	\includegraphics[width=2\columnwidth]{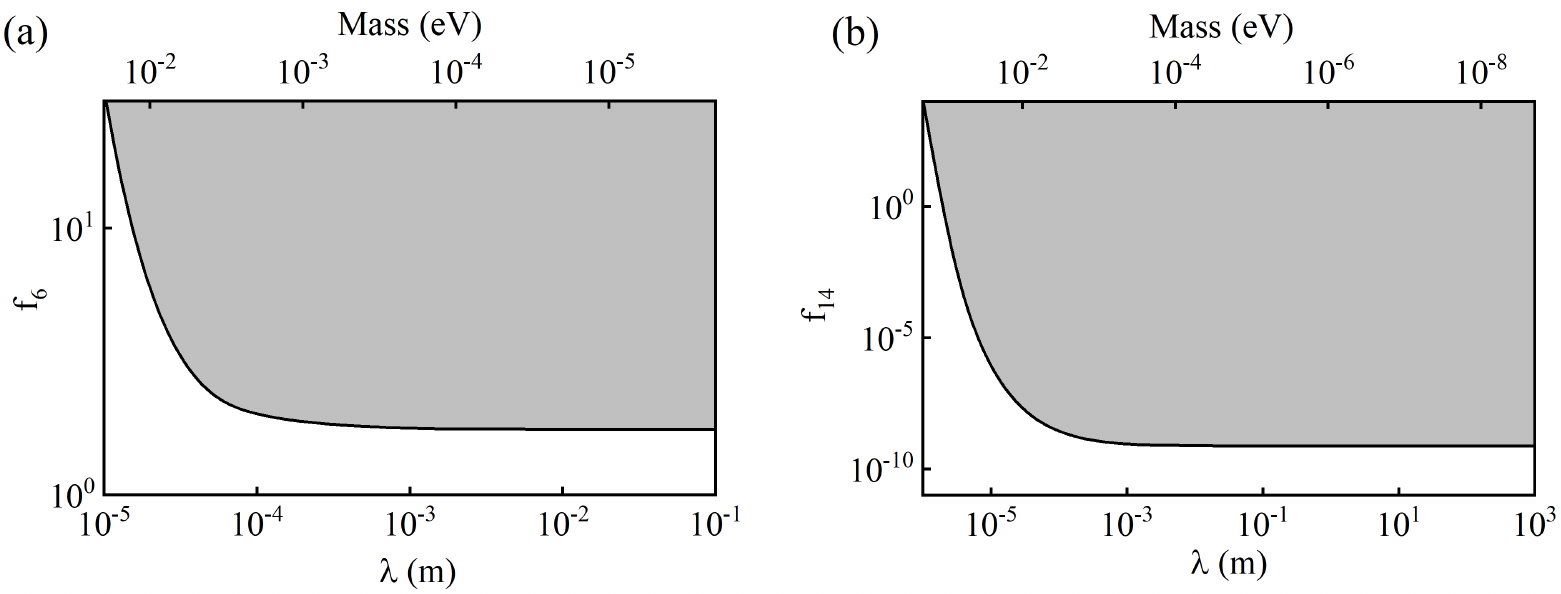}
	\caption{(a) Experimental constraints on $f_{6}$ between electrons as a function of the force range $\lambda$ and the boson mass. The gray filled regions are excluded parameter spaces. The black line indicates upper limit on the coupling established by our experiment for the force range $\lambda<1 $ cm. (b) Experimental constraints on $f_{14}$ between electrons. Our work sets new constraints for the force range $\lambda<1 $ km. }
	\label{SAbound}
\end{figure*}

The total searching experiment was performed for 26 hours to reduce statistical uncertainty. The mean value and the standard error of each data-set are shown in Fig. \ref{SAresult}(c), where the fit in the inset indicates that each set of data follows a Gaussian distribution. With the overall 26-hour data, the first order amplitude of the effective magnetic field $B^{(1)}_{\text{eff}}$ is determined to be ($-2.8\pm7.8$) pT with the reduced chi-square $\chi^2$ = 0.82.
The mean value of the measured effective field is smaller than its standard deviation, indicating no evidence of exotic SSVDIs in this experiment. This sets new limits on the coupling coefficients corresponding to SSVDIs.

Systematic errors are summarized in Table \ref{t3}, where we take $\lambda=1$ mm as an example. The main contributions come from the uncertainties of geometric parameters of the spin source, such as the radius of the polarized NV area, the thickness of the NV layer, and the deviation of the spin source and the spin sensor in $x$-axis direction. The polarized spin density $\rho_{\rm pol}$ and its uncertainty were obtained by monitoring the magnetic dipole interaction over a long period. Other systematic errors include the uncertainty of the phase $\phi$ of demodulation reference signal, the fluctuation of coefficient $\eta$ between the magnetometer output voltage signal and the sensed magnetic field. We also analyzed some other possible sources of systematic errors which are not listed in the table due to their negligible effect, such as the effects of the moving surface charges and the demagnetization factor of the spin source (see Appendix \ref{SM4} for details). The overall systematic error was derived by assuming the systematic uncertainties independent of each other and combining all of them in quadrature. Therefore, we quote the final coupling coefficient as  $f_{6}=(-0.27\pm 0.76_{\rm stat} \pm0.06_{\rm sys} )$ for $\lambda =$1~mm, which determines $|f_{6}|<1.76$ at the 95$\%$ confidence level. By varying the force range and repeating a similar procedure, the constraints on coupling coefficients for the explored force ranges can be obtained. It is important to note that the geometric factors for different force ranges have been accounted for in the calculation.

Figure \ref{SAbound}(a) shows the experimental constraints on $f_{6}$ established by our work. For the force range from 1 cm to 1 km, the most stringent constraints were set by Ji \textit{et al.} \cite{Ji2018RPL}, in which a spin-exchange-relaxation-free (SERF) comagnetometer was utilized to detect the possible effective magnetic field created by rotating SmCo$_5$ permanent magnets as electron spin sources. Our experiment, benefiting from the small size of ensemble-NV-diamonds, is more sensitive to exotic interactions at micrometer scale and sets stringent limits in the force range from around 10 $\upmu$m to 1 cm. 

Furthermore, our results can also be utilized to constrain another SSVDI  between electrons:
\begin{equation}
	V_{14}=f_{14}\frac{\hbar}{4\pi}[\hat{\sigma_1}~\times~\hat{\sigma_2})\cdot\vec{v}]\left(\frac{1}{r}\right)e^{-r/\lambda}.
	\label{V14}
\end{equation}
Only one preceding experiment has constrained directly on $f_{14}$ using electron spins within the earth as the spin source\cite{Hunter2014}. However, that work has not provided constraints for $\lambda<1$ km, where fluctuations in the local polarized geoelectron density and potential local ferromagnetic interference will render the results unreliable at short ranges\cite{Hunter2013}. As shown in Fig. \ref{SAbound}(b), our work explores the parameter space inaccessible for the geoelectron experiment, and offers new direct constraints on $f_{14}$ in the force range of 1 $\upmu$m to 1 km.

\begin{table}[h!]
	\caption{Summary of systematic errors. The corrections to  $f_{6}$ and $f_{14}$ with $\lambda=1$~mm are listed.}
	\label{t3}
	\resizebox{\linewidth}{22mm}{
		\begin{tabular}{l c c c}
			\hline
			\hline
			Parameter&Value&$\Delta f_{6}$  & $\Delta f_{14} (10^{-10})$\\
			\hline
			Diameter 2$R$ & $ 104\pm 1\ \upmu$m&$\pm0.01$ & $\pm0.03$\\
			Thickness $h_1$& $23 \pm 1 \upmu$m &$\pm0.01$& $\pm0.01$\\
			Thickness $h_2$& $23 \pm 1 \upmu$m &$\pm0.02$& $\pm0.06$\\
			Amplitude $A$& $36.7\pm 0.5$ nm & $\pm0.01$ & $\pm0.02$\\
			Distance $d$& $18.5\pm 0.6\ \upmu$m &$\pm0.01$ & $\pm0.01$\\
			Phase $\phi$& $-6.7\pm 4.4^\circ$ &$\pm0.01$ & $_{-0.02}^{+0.01}$\\
			Deviation in x & $46\pm 1\ \upmu$m &$_{-0.05}^{+0.04}$&  $\pm0.01$\\
			Relative angle & $54.0 \pm 0.4^\circ $ &$\pm 0.01$&  $\pm 0.01$\\
			Coefficient $\eta$ &$4.1\pm0.1$~V/mT& $\pm0.01$& $\pm0.03$\\
			Polarized Density~$\rho_{\rm pol}$  &$(1.14\pm0.02)\times10^{23}~$m$^{-3}$& $\pm0.01$& $\pm0.02$\\
			\hline
			\multirow{2}{2 cm}{Final $f_{6} $} &\multirow{2}{1.9 cm}{$-0.27$}&$\pm 0.76$ (statistic)& \\
			& &$\pm 0.06$ (systematic)& \\
			\hline
			\multirow{2}{2 cm}{Final $f_{14} $} &\multirow{2}{1.9 cm}{$-1.36\times10^{-10}$}&&$\pm 3.80$ (statistic) \\
			& && $\pm 0.08$ (systematic)\\
			\hline
			\hline
	\end{tabular}}
\end{table}

In summary, we report a new experimental search of two types of SSVDIs between polarized electrons. Using an NV ensemble as the spin sensor and another high-concentration NV ensemble as the spin source, we set new limits on $V_6$ and $V_{14}$ at the micrometer scale. We anticipate that further advances in experimentation will facilitate the search process in the future. To achieve higher polarized spin density, we can use high-power laser and microwave pulses to polarize the spin source. Moreover, using a silicon carbide heat spreader connected to the diamond can mitigates laser-induced thermal effect\cite{schloss_simultaneous_2018}, and employing an infrared absorption readout method can effectively improve the detection efficiency\cite{chatzidrosos_miniature_2017}. The application of the pulsed magnetic detection method is expected to achieve a better signal contrast. Therefore, the magnetic sensitivity of the ensemble-NV-diamond  magnetometer can be improved in the future. We note that despite the fact that other SSVDIs like $V_{7}$, $V_{15}$, $V_{16}$ vanish between two identical electrons as a result of commutative antisymmetry\cite{Fadeev2019a,dobrescu2006spin}, these forms of exotic interactions involving electron spins and polarized nucleons may still exist and allow detection utilizing the extension of our platform. With the development of spin-mechanical quantum chip technology, the exotic interactions can be investigated at shorter force range\cite{wu_spin-mechanical_2023}. Overall, taking advantages of manipulation of the polarized spin states, NV ensembles have demonstrated a potential for searching exotic spin-spin interactions beyond the Standard Model.

This work was supported by the Innovation Program for Quantum Science and Technology (2021ZD0302200), the Chinese Academy of Sciences (No. GJJSTD20200001), the National Key R\&D Program of China (Grant No. 2018YFA0306600, No. 2021YFC2203100), Anhui Initiative in Quantum Information Technologies (Grant No. AHY050000), NSFC (12150010, 12205290, 12261160569, 12261131497). X. R. thanks the Youth Innovation Promotion Association of Chinese Academy of Sciences for the support. Y. F. C., Y. W., M. J. and Y. X. thank the Fundamental Research Funds for Central Universities. Y. F. C. is supported in part by CAS Young Interdisciplinary Innovation Team (JCTD-2022-20), 111 Project (B23042). M. J. is supported in part by China Postdoctoral Science Foundation (2022TQ0330). This work was partially carried out at the USTC Center for Micro and Nanoscale Research and Fabrication.

Y. H. and H. L. contributed equally to this work.

\appendix

\section{Experimental setup}
\label{SM1}
\subsection{Schematic of the experimental setup}

\begin{figure}[htbp]
	\centering
	\includegraphics[width=1\columnwidth]{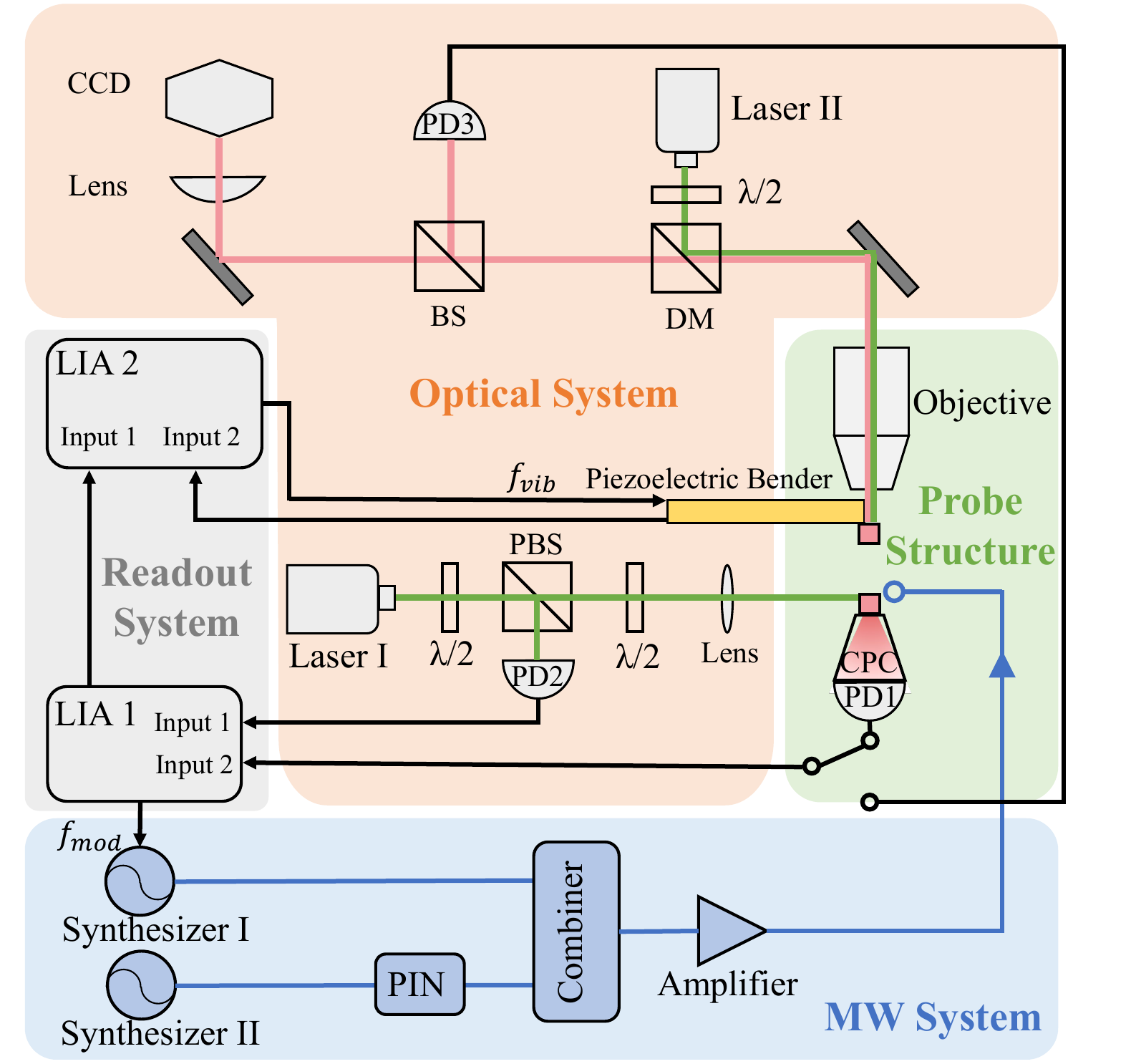}
	\caption{\textbf{The schematic of the experimental setup.} $\lambda$/2, halfwave plate; BS, beam splitter; PBS, polarizing beam splitter; DM, Dichroic Mirror; CPC, compound
		parabolic concentrator; PD, photodiode; LIA, lock-in amplifier; PIN, positive-intrinsic-negative diode. $f_{\rm mod}$, the modulation frequency of the microwave of the spin sensor; $f_{\rm vib}$, the vibration frequency of the spin source. }
	\label{SMsetup}
\end{figure}

The scheme of the experimental setup is shown in Fig.\ref{SMsetup}. The spin sensor was a 23-$\upmu$m-thick nitrogen-vacancy (NV) ensemble at the surface of diamond \uppercase\expandafter{\romannumeral1} with a concentration being 14 ppm. The 532-nm laser (Laser I) provided by a high-power optically pumped semiconductor laser (Combolt, 0532-05-01-1500-700) was focused by a 5-cm lens and illuminated the spin sensor via the flank of the diamond. A half-wave plate was utilized to adjust the optical polarization of the laser. The red fluorescence emitted from the spin sensor was collected through a compound parabolic concentrator, filtered by a long-pass filter and detected by a photodetector (PD1, Thorlabs, SM05PD1A). Another PD (PD2, Thorlabs, SM05PD1A) was utilized to record the power fluctuation of Laser I for noise cancellation.

The spin source was the other 23-$\upmu$m-thick NV layer at the bottom of diamond \uppercase\expandafter{\romannumeral2} with a concentration of NV centers being 22 ppm. The spin source was attached to a piezoelectric bender and was placed above the spin sensor where the distance between them was $d = 18.5~\upmu$m. The green laser (Laser II) was sent through an objective (Nikon, CFI S Plan Fluor ELWD 20XC) above the spin source. The fluorescence from the spin source was collected by the same objective and then detected by a photodetector (PD3, Thorlabs, SM05PD1A). A 500-nm-silver layer was coated on the bottom of diamond \uppercase\expandafter{\romannumeral2} to isolate the two laser beams as well as fluorescence. A charge-coupled device (CCD, Thorlabs, CS165MU) was used to monitor and estimate the effective areas of the spin sensor and the spin source.

A 94 gauss static bias magnetic field $B_0$ was applied along one of the NV axes of the spin sensor with a permanent magnet. The two pieces of diamonds were carefully positioned with a relative rotation angle of 54$^\circ$ along the vertical axis so that the projection of $B_0$ onto the NV axis differed for the spin sensor and the spin source. Therefore, the resonant frequencies of the microwave applied on the spin source and the spin sensor were different. The resonant microwave applied on the spin sensor (source), MW I (II), with frequency being 3106 (3040) MHz was generated by Synthesizer I (II) (National Instrument, FSW-0010). Microwave fields from two microwave channels were combined by a combiner (Mini-circuits, ZN2PD2-14W-S+), amplified by a power amplifier (Mini-circuits, ZHL-15W-43-S+) and delivered to both the spin source and the spin sensor via a double-split-ring-resonator.

MW I was modulated with the modulation frequency $f_{\rm mod} $ = 79.426~kHz. The signal of PD1 and PD2 were demodulated by LIA1 (Zurich Instruments, HF2LI) with the frequency $f_{\rm mod}$. Another lock-in amplifier (LIA2, Zurich Instruments, HF2LI) was used to drive the vibration of the piezoelectric bender at frequency $f_{\rm vib}$ = 1.337 ~kHz and demodulate the magnetometer signal from LIA1 with $f_{\rm vib}$.

\begin{figure}[htbp]
	\centering
	\includegraphics[width=1\columnwidth]{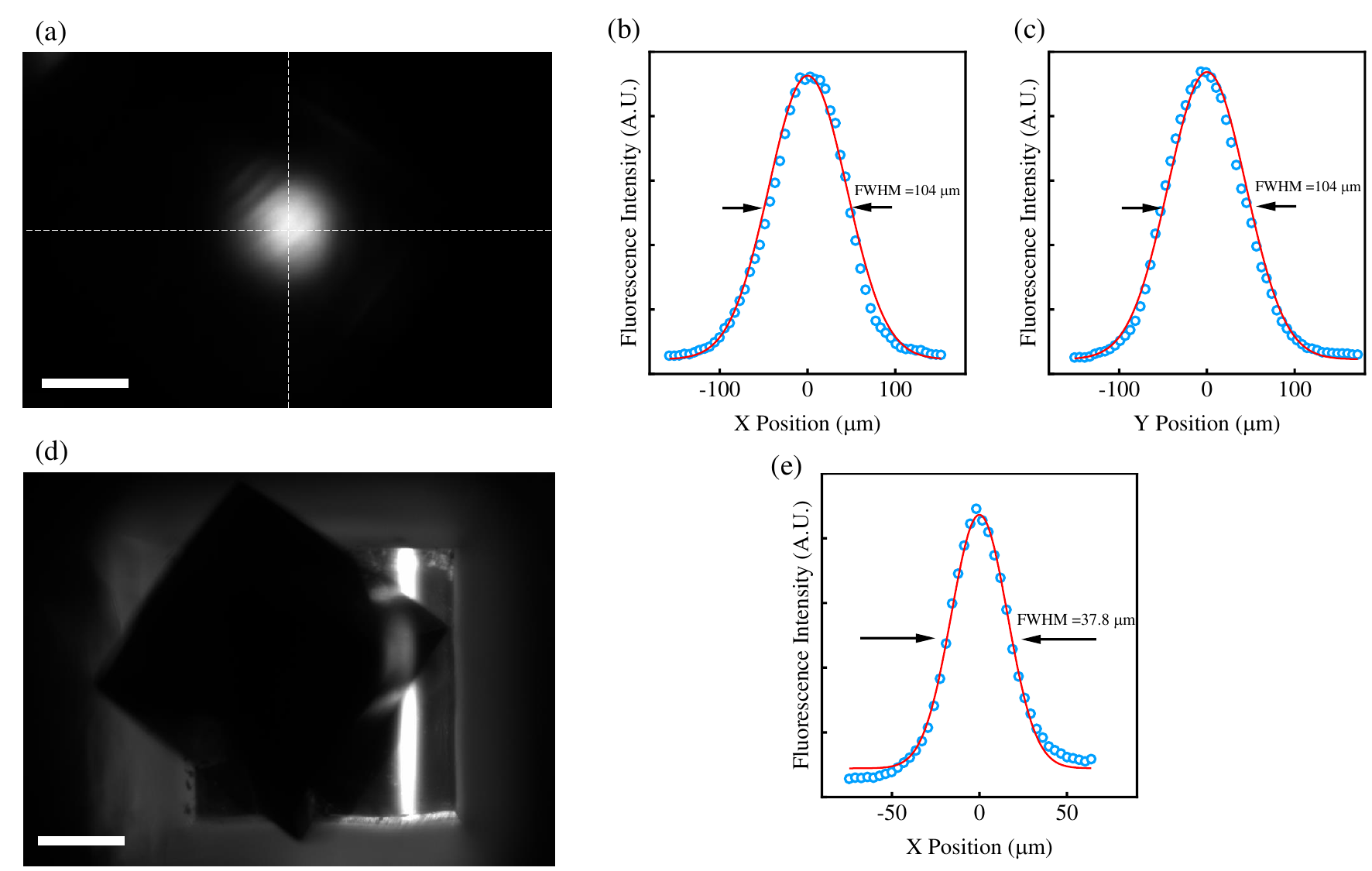}
	\caption{\textbf{Measurement of the area of the spin sensor and the spin source.} (a) Fluorescence image of the spin source NV volume. (b) (c) Slides through the image in (a) (horizontal, (b); vertical, (c)) fitted to Gaussian line shapes. The FWHM of the spot measured by CCD was 104 $\upmu$m. (d) Fluorescence image of the spin sensor NV volume. (e) Horizontal cut through the image in (d) fitted to a Gaussian line shape.}
	\label{SMspot}
\end{figure}

\subsection{Estimation of the area of the spin sensor and the spin source}

The areas of the spin sensor and the spin source were measured by fluorescence images with a CCD, with the method introduced in  Ref.\cite{glenn_2018}. Figure \ref{SMspot}(a) shows the fluorescence image of the spin source. Figure \ref{SMspot}(b) and (c) show the horizontal and vertical slices of the image and the Gaussian fits to the lines.
The extracted spot size is 104(1) $\upmu$m FWHM both horizontally and vertically. Thus the spin source volume is modeled as a cylinder with a radius $R$ = 52 $\upmu$m and a height of 23 $\upmu$m. Figure \ref{SMspot}(d) shows the fluorescence image of the sensing volume of the spin sensor. The laser was applied from the flank of the diamond and formed a long strip on the NV layer of the diamond. Part of the light spot was blocked by the rotated diamond above, whose bottom was coated with silver. Figure \ref{SMspot}(e) shows the horizontal slice through the unblocked part and a fitting with a Gaussian line being 37.8 $\upmu$m FWHM. Thus the volume of the spin sensor can be taken as a cuboid of $(37.8\times660\times23)~ \upmu$m$^3$.

\begin{figure}[htbp]
	\centering
	\includegraphics[width=1.0\columnwidth]{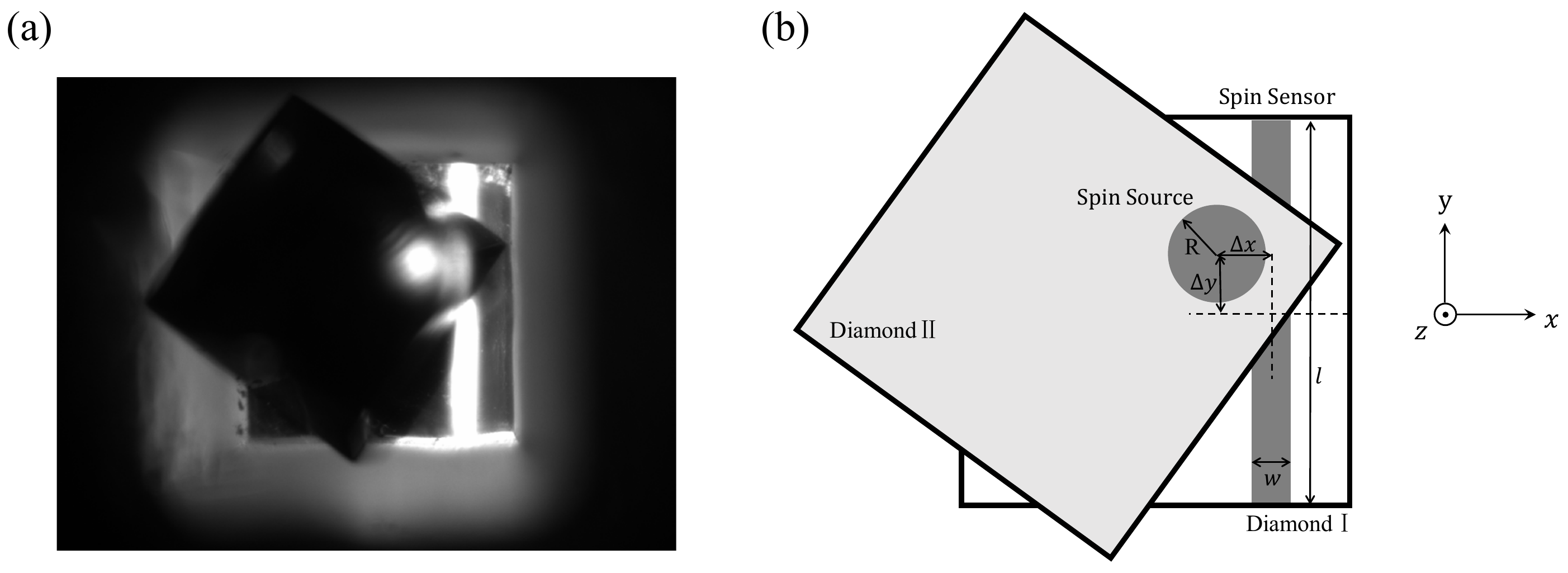}
	\caption{(a) Fluorescence image of the spin sensor together with the spin source. (b) Schematic diagram of experimental parameters from a top view.} 
	\label{SMparameter}
\end{figure}

Figure \ref{SMparameter}(a) is the fluorescence image of the spin sensor and the spin source, and how they are positioned relative to each other. Figure \ref{SMparameter}(b) shows a top view schematic diagram. The sensing area of the spin sensor is a long strip with a length $l = 660~ \upmu$m, width $w = 37.8~ \upmu$m and height $h_1 = 23~ \upmu$m. The area of the spin source is a cylinder with radius $R = 52~\upmu$m and height $h_2 = 23~ \upmu$m. The centers of the spin source and the spin sensor were offset by $\Delta x = 46~ \upmu$m in the $x$ direction and $\Delta y = 90~ \upmu$m in the $y$ direction.

\subsection{Distinct resonant frequencies of the spin sensor and the spin source}

\begin{figure}[htbp]
	\centering
	\includegraphics[width=0.8\columnwidth]{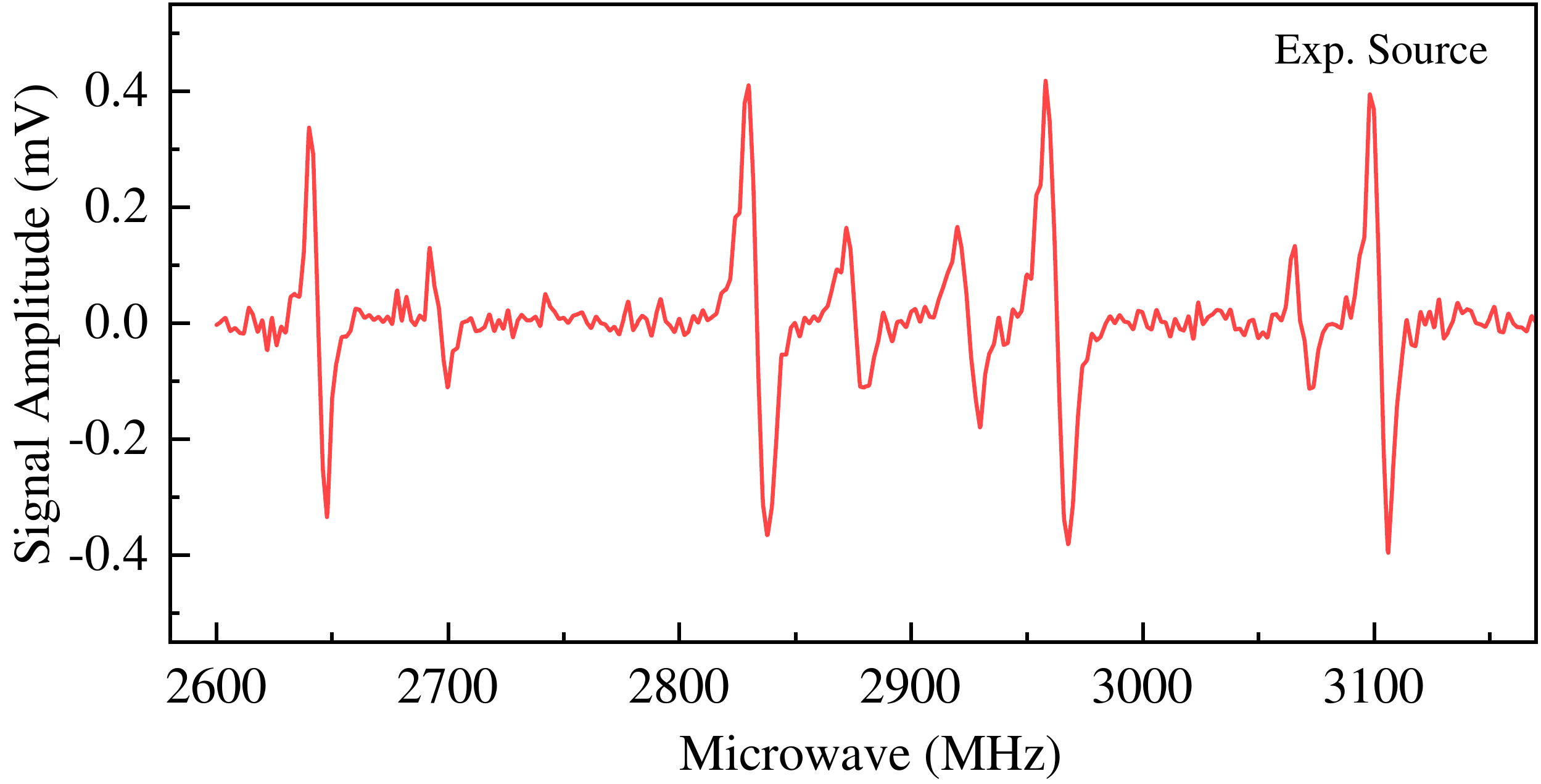}
	\caption{\textbf{The experimental ODMR spectrum of the spin source.} The eight resonance correspond to the $|m_s=0\rangle \to |m_s=+1\rangle$ and $|m_s=0\rangle \to |m_s=-1\rangle$  transitions of the four NV axes, respectively. }
	\label{SMCW_up}
\end{figure}

We positioned the spin sensor and the spin source with a relative angle so that the NV centers of sensor and source can be manipulated independently by microwaves with two different frequencies. Figure \ref{SMCW_up} shows the first-order differential Optically Detected Magnetic Resonance (ODMR) spectrum of the source with a 94 gauss static field $B_0$ along the sensor's NV axis. The power of Laser II was set to be 100 mW. Note that Laser I was turned off during this measurement. The eight resonance signals correspond to the $|m_s=0\rangle \to |m_s=+1\rangle$ and $|m_s=0\rangle \to |m_s=-1\rangle$  transitions of the four NV axes.

\begin{figure}[htbp]
	\centering
	\includegraphics[width=0.8\columnwidth]{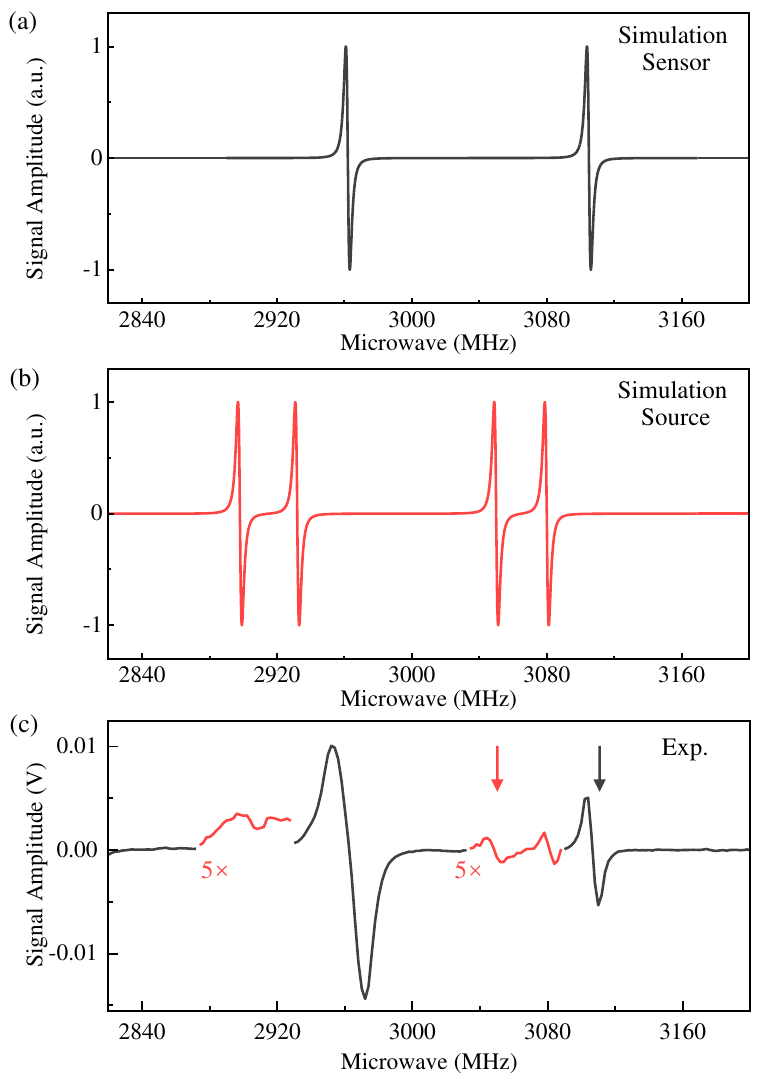}
	\caption{\textbf{} (a) The simulated ODMR spectrum of the spin source in magnetic field $B_0$. (b) The simulated ODMR spectrum of the spin sensor. (c) The experimental ODMR spectrum of the spin sensor and the spin source. Part of the data (red lines) have been magnified by 5 times. The red and black arrows point to the two microwave frequencies applied in the experiment to manipulate the spin source and the spin sensor, respectively.}
	\label{SMCW_updown}
\end{figure}

When we turned on both lasers, the simulated and experimental first-order differential ODMR spectrum of the spin source and spin sensor are shown in Fig. \ref{SMCW_updown}(a) and (b), respectively. Only $|m_s=0\rangle \to |m_s=+1\rangle$ transitions are presented here. 
Figure \ref{SMCW_updown}(c) was obtained by demodulating the fluorescence from the spin sensor and the spin source, with Laser I of 600 mW and Laser II of 100 mW. Part of the data has been magnified by 5 times, because the fluorescence from the spin source was much weaker than that of the spin sensor. In addition, due to the heating effect of Laser, the zero-field splitting in Fig. \ref{SMCW_updown} was shifted by 22 MHz compared with that in Fig. \ref{SMCW_up}. In our experiment, we applied resonant microwave field to control the spin states of the spin source (sensor), as indicated by the red (black) arrow in Fig. \ref{SMCW_updown}(c). For the spin source, the frequency of the resonant microwave was 3040 MHz. For the spin sensor, the frequency of the resonant microwave was 3106 MHz.

\subsection{The performance of the  ensemble-NV-diamond magnetometer}

\begin{figure}[htbp]
	\centering
	\includegraphics[width=0.8\columnwidth]{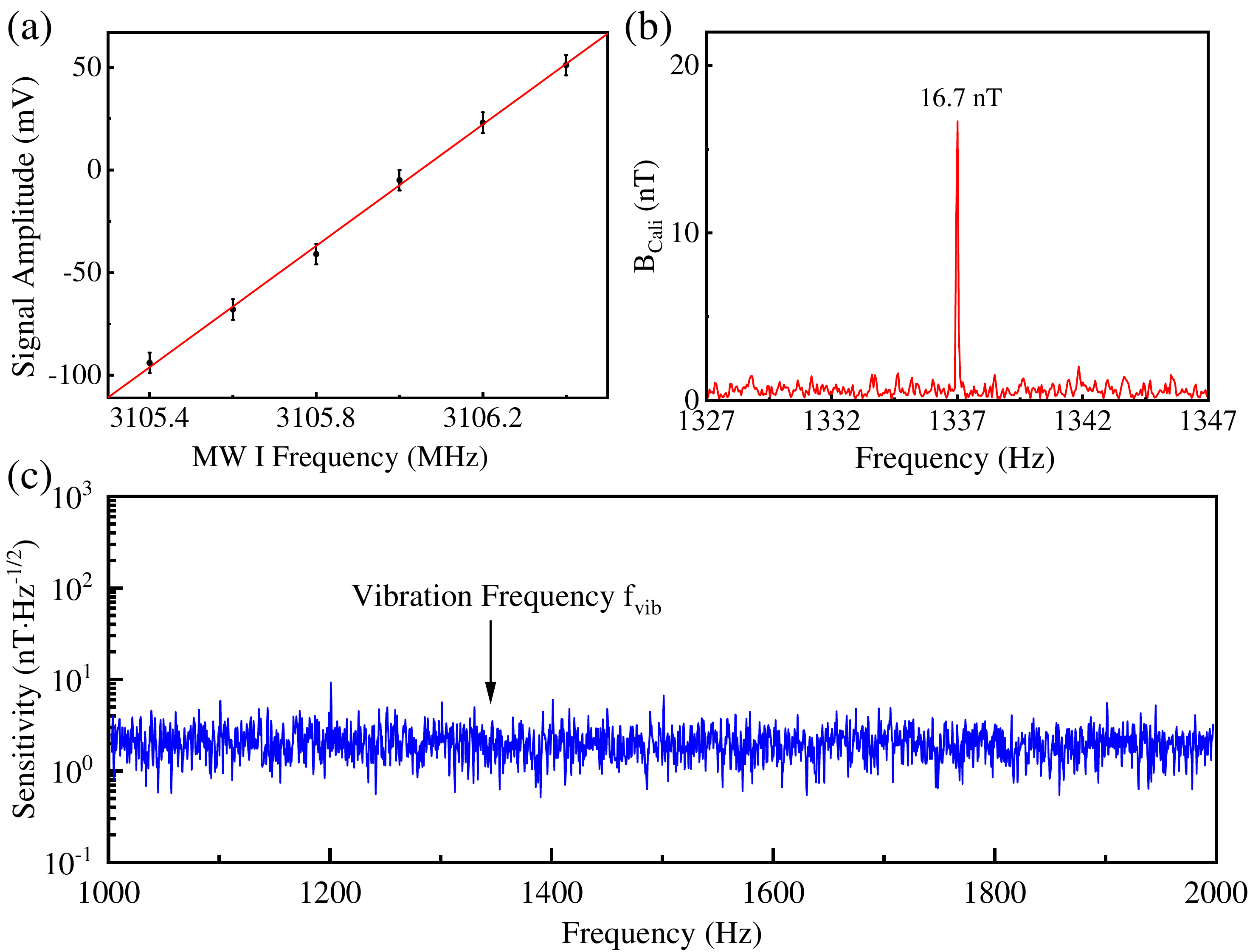}
	\caption{(a) The specific region of the CW spectrum of the NV magnetometer. The red line is the linear fitting to obtain the max slope. (b) Verification of magnetic sensitivity at 1337 Hz using a copper coil. (c) Magnetic sensitivity of the NV ensemble magnetometer with the frequency of 1 to 2 kHz. The vibration frequency of the spin source $f_{\rm vib}$ = 1.337~kHz is also displayed.}
	\label{SMsensi}
\end{figure}

The spin sensor is an NV ensemble magnetometer with continuous-wave(CW) method. The excitation laser and the resonance microwave field were continuously applied on the spin sensor, and the fluorescence was detected continuously. The magnetometer output depended on the magnetic field with a coefficient, $\eta$, was determined by the max slope of the CW spectrum, which was 148(3) mV/MHz as shown in Fig. \ref{SMsensi}(a). This corresponded to a coefficient $\eta$ of $4.1 \pm 0.1  $~V/mT. The magnetic sensitivity of 2 nT/Hz$^{1/2}$ from 1 to 2 kHz was achieved, as shown in Fig. \ref{SMsensi}(c).

The sensitivity of the NV ensemble magnetometer at the target frequency $f_{\text{vib}}$ was also verified. A magnetic field along the $x$-axis with frequency of 1337 Hz was applied by a copper coil, and was measured to be 20.1 nT in $x$-axis by a commercial flux gate (HSF113-2H6-AAA). With the NV magnetometer, the magnetic field along the NV axis ($-\sqrt{2/3}\hat{x},~0,~\sqrt{1/3}\hat{z}$) was measured to be 16.7 nT, as shown in Fig. \ref{SMsensi}(b). It corresponded to a 20.4 nT magnetic field in the $x$-axis direction, which was in good agreement with the result of the commercial flux gate. Thus, the magnetic sensitivity of 2 nT/Hz$^{1/2}$ at $f_{\text{vib}}$ was verified.


\section{Measurement of the polarized spin density}
\label{SM2}

\subsection{Modulation of the polarization of the spin source}

\begin{figure}[htbp]
	\centering
	\includegraphics[width=0.7\columnwidth]{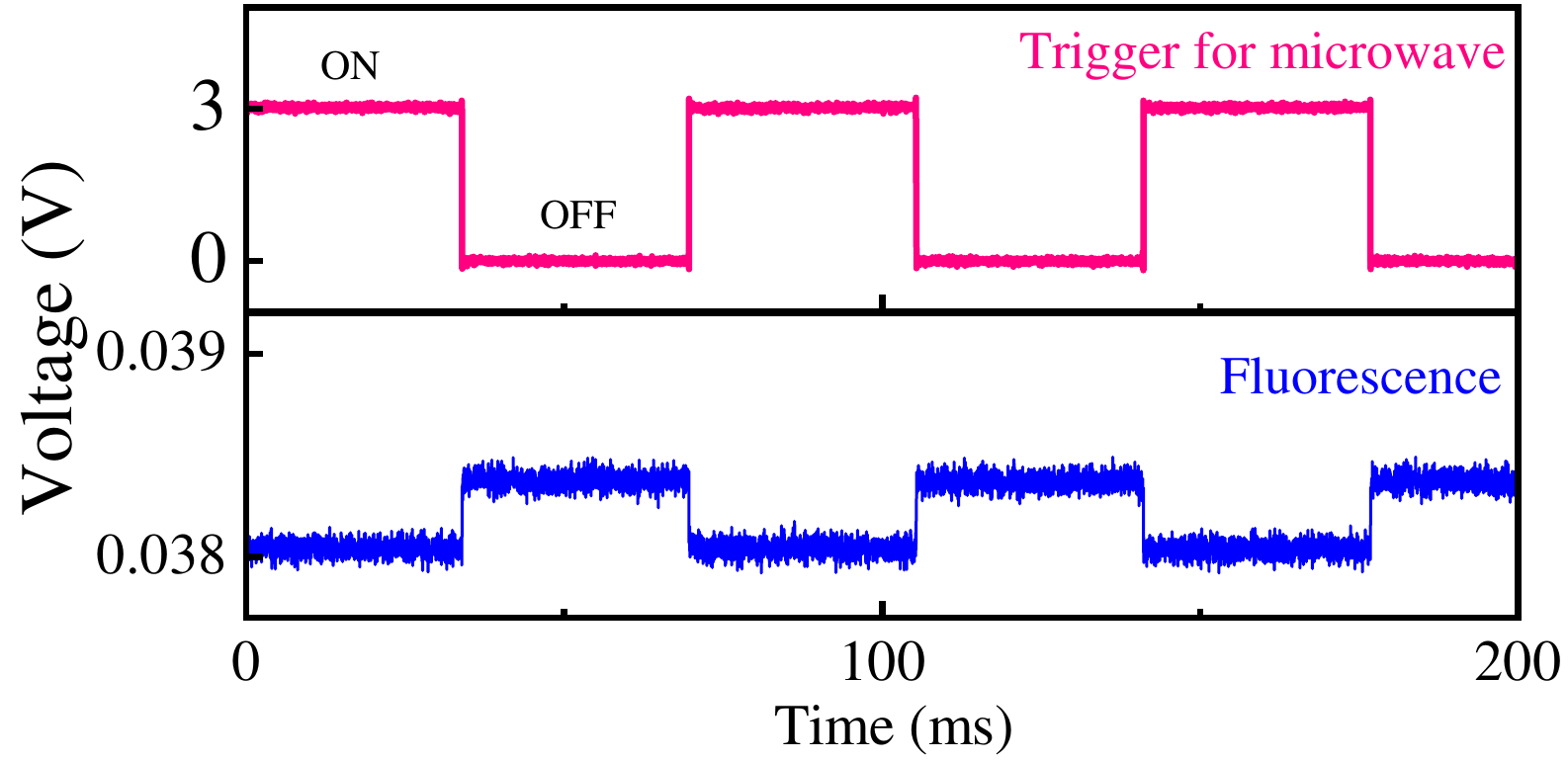}
	\caption{ Variation of fluorescence intensity of the spin source with the microwave switched on and off periodically.}
	\label{SMFL}
\end{figure}

The magnetic dipole-dipole interaction between the spin sensor and the spin source was measured to obtain the magnetic polarized spin density $\rho_{\rm pol}$ of the spin source before searching for exotic SSVIDs.
To measure the magnetic polarized spin density, we modulated the polarization of the spin source by switching MW II on and off periodically. An arbitrary waveform generator (AWG, Rigol DG812) generated a Transistor Transistor Logic (TTL) sequence to control the microwave switch through a PIN. By monitoring the intensities of fluorescence, we were able to determine the changes in the polarization state of the spin source. The upper objective collected the fluorescence from the spin source and filtered it before sending it to a PD for detection. The TTL sequence and the signal of PD were simultaneously recorded by an oscilloscope.
The experimental result of the variation of fluorescence intensity with a time duration of 200 ms is shown in Fig. \ref{SMFL}. The upper panel shows that the microwave was switched on and off periodically with a 50\% duty cycle at 14 Hz. When the voltage of the TTL was set to a high level, the resonant microwave field was applied and the population of the $\ket{m=0}$ state was partially transferred to $\ket{m=+1}$ state, resulting in a reduction of fluorescence intensity, as illustrated in the lower panel. It indicates that the polarization changes quickly during the switching process of the microwave. The spin source becomes magnetic polarized when the microwave was switched on, and this effect vanished when the microwave was turned off.

\subsection{Numerical simulation of the magnetic field due to the magnetic dipole-dipole interaction}
\label{Numerical calculations of the magnetic dipole fields}
The magnetic dipole field can be expressed as,

\begin{equation}
	B_{\rm dipole} = -\frac{\mu_0 \gamma \hbar}{8\pi}\frac{3(\hat{\sigma_1} \cdot \hat{r})( \hat{\sigma_2} \cdot \hat{r})-(\hat{\sigma_2} \cdot \hat{\sigma_2})}{r^3}
	\label{B_dipole}
\end{equation}
where $\mu_0$ is the vacuum permeability, $\gamma$ is the gyromagnetic ratio, $\hat{\sigma_2}$ is the unit spin vector of a single electron spin in the spin source, $\hat{r}$ is the unit displacement vector.

By integrating the volume of both the spin source and sensor, we derive the magnetic field sensed by the spin sensor due to the magnetic dipole-dipole interaction as,
\begin{equation}
	B_{\rm d} = -\frac{\mu_0 \gamma \hbar}{8\pi}\frac{1}{V_{\rm\uppercase\expandafter{\romannumeral1}}}  \rho_{\rm pol}\int_{V_{\rm\uppercase\expandafter{\romannumeral1}}} dV\int_{V_{\rm\uppercase\expandafter{\romannumeral2}}}dV' \frac{3(\hat{\sigma_1} \cdot \hat{r})( \hat{\sigma_2} \cdot \hat{r})-(\hat{\sigma_2} \cdot \hat{\sigma_2})}{r^3}
	\label{B_dmain}
\end{equation}
where $V_{\rm\uppercase\expandafter{\romannumeral1}}(V_{\rm\uppercase\expandafter{\romannumeral2}})$ stands for the integration volume of the spin sensor (spin source), $\rho_{\rm pol}$ is the polarized spin density of the spin source.

The Monte Carlo method was utilized to numerically calculate $B_{\rm d}$, as same as in Ref. \cite{liang_new_2022}. The algorithm of the Monte Carlo integral is performed as follows:
\begin{itemize}
	\item[(1)] $N_{\rm MC}=2^{20}$ random pairs of points inside both the volumes of the spin sensor and the spin source are generated.
	\item[(2)] The magnetic field $B_{\rm dipole}^i$ between a randomly generated pair of points is calculated using Eq. (\ref{B_dipole}).
	\item[(3)] All the contributions to the magnetic fields are summed and normalized to give the average magnetic field generated by the spin source and sensed by the spin sensor:
	\begin{equation}
		B_{\rm d} = \rho_{\rm pol}V_{\rm\uppercase\expandafter{\romannumeral2}}\frac{1}{N_{\rm MC}}\sum_{i}^{N_{\rm MC}} B_{\rm dipole}^i
		\label{B_d}
	\end{equation}
\end{itemize}

\begin{figure}[htbp]
	\centering
	\includegraphics[width=1\columnwidth]{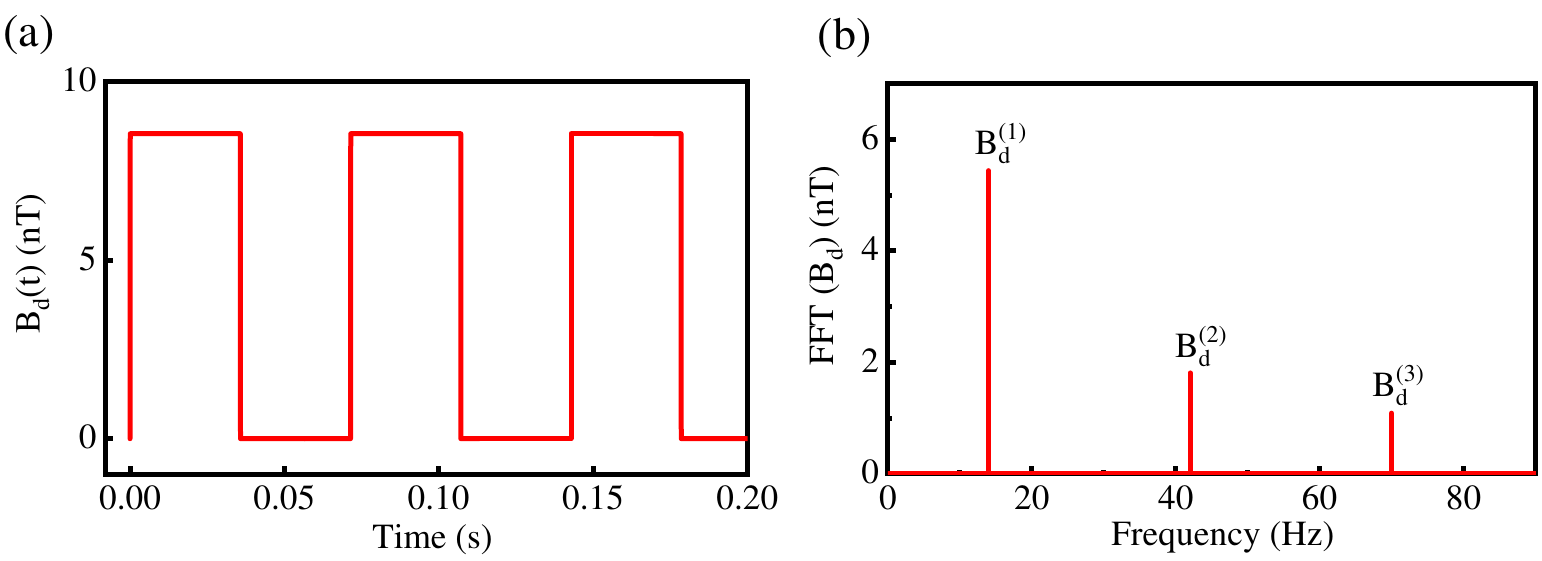}
	\caption{\textbf{Numerical simulation of the magnetic field $B_{\rm d}$. } With a periodically switched microwave MW \uppercase\expandafter{\romannumeral2}, the spin source generates a square-wave magnetic dipole field $B_{\rm d}$ on the spin sensor with 50\% duty cycle at a frequency $f_{\rm d}$ = 14 Hz. Here we take the polarized spin density $\rho_{\rm pol} = 9.68\times 10^{22}~m^{-3}$ as an example to numerically calculate the magnetic dipole field $B_{\rm d}$. (a) Time-domain signal of $B_{\rm d}$. (b) Fourier transformation spectrum of $B_{\rm d}$.}
	\label{SMdipolecal}
\end{figure}

As discussed above, the polarization of the spin source is modulated periodically with a 50\% duty cycle and a frequency of 14 Hz. Correspondingly, the magnetic field due to the magnetic dipole-dipole interaction can be characterized as a square wave with frequency $f_{\rm,d}$ = 14 Hz. The modulated magnetic field can be decomposed into a series of odd sinusoidal harmonics:
\begin{equation}
	B_{\rm d}(t) = \sum_{\rm n=odd}^\infty B_{\rm d}^{(n)}\sin(2\pi nf_{\rm d}t),
	\label{B_dfft}
\end{equation}
the coefficient is derived as $B_{\rm d}^{(n)} = \frac{2}{T}\int_0^T\sin(2\pi nf_{\rm d}t)B_{\rm d}(t)dt$, where $T = 1/f_{\rm d}$ is the period of the modulation. First, we perform numerical simulation to demonstrate the resulted magnetic field $B_{\rm d}$ on the senor due to the magnetic dipole-dipole interaction. The total density of NV centers in the spin source is 3.87$\times 10^{24}~$m$^{-3}$~(22~ppm). As an example, here we assume that 10\% of the NV centers are polarized in the spin source.  Only one of four symmetry axes of NV centers can be polarized by the resonant microwave. The polarized spin density is $\rho_{\rm pol} = 3.87\times 10^{24}~$m$^{-3}\times\frac{1}{4}\times10\% = 9.68\times 10^{22}~$m$^{-3}$. Figure \ref{SMdipolecal} shows the simulated magnetic field due to the magnetic dipole-dipole interaction, $B_{\rm d}$. In the time domain, $B_{\rm d}$ varies from 0 to 8.55 nT.
In our experiment, we focus on the first-order harmonic of the Fourier transformation spectrum $B_{\rm d}^{(1)}$, whose amplitude is estimated to be 5.44 nT. If we obtained $B_{\rm d}^{(1)}$ from the lock-in amplifier, the polarization spin density of the spin source can be obtained experimentally.

\subsection{Experimentally measuring the magnetic field due to the dipole-dipole interaction to determine the magnetic polarized spin density}

\begin{figure}[htbp]
	\centering
	\includegraphics[width=1\columnwidth]{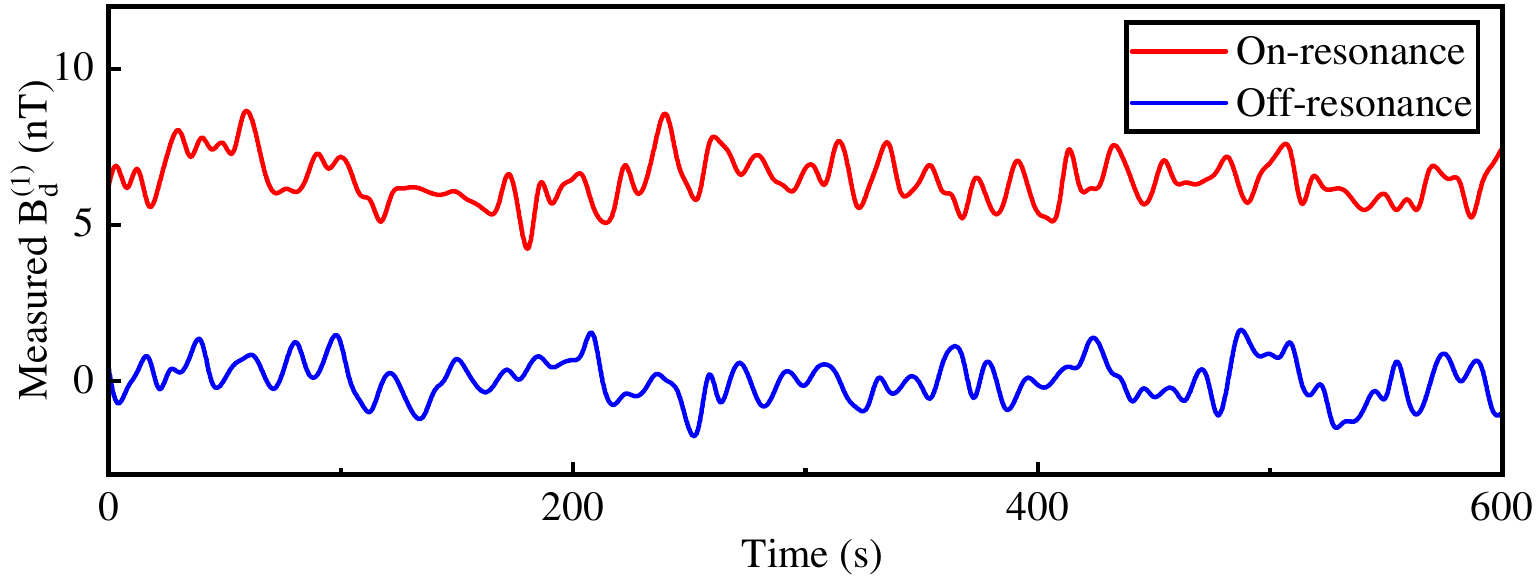}
	\caption{\textbf{Measurement results of $B_{\rm d}^{(1)}$ under on-resonance and off-resonance conditions.} Red and blue lines are the experimental magnetic fields $B_{\rm d}^{(1)}$ with the on and off resonance microwave, respectively.}
	\label{SMdipolemeasure}
\end{figure}

The detection scheme has been shown in Fig. 2 in the main text. We extract the amplitude of the first-order harmonic $B_{\rm d}^{(1)}$ by demodulating the ensemble-NV-diamond magnetometer with frequency $f_{\text{d}}$ using LIA2 in Fig. \ref{SMsetup}. The measured results of $B_{\rm d}^{(1)}$ are shown in Fig. \ref{SMdipolemeasure} with a time duration of 600~s. When MW \uppercase\expandafter{\romannumeral2} was on-resonance and switched on and off periodically with a 50\% duty cycle at a frequency of 14 Hz, $B_{\rm d}^{(1)}$ was measured to be 6.41 $\pm$ 0.09~nT. We also carried out a controlled experiment when MW \uppercase\expandafter{\romannumeral2} was off-resonance and the spin source were not magnetic polarized. The experimental value of $B_{\rm d}^{(1)}$ in this case was 0.04 $\pm$ 0.08~nT. Given the numerical simulation results from the previous text, the polarized spin density of the spin source can be obtained as $\rho_{\rm pol} = 1.14(2)\times10^{23}~$m$^{-3}$. Note that, during the measurement of the polarized spin density, the spin source remained stationary and the SSVDIs is zero.

\section{Search for SSVDIs}
\label{SM3}
\subsection{Numerical simulations of the effect of SSVDIs}
In this section, we present numerical simulations of the effective magnetic fields generated by the SSVDIs. The effective magnetic fields can be expressed as,
\begin{equation}
	B_6 = -f_6\frac{\hbar}{2\pi m_e\gamma c}[(\hat{\sigma_1}\cdot\vec{v})(\hat{\sigma_2}\cdot\hat{r})+(\hat{\sigma_1}\cdot\hat{r})(\hat{\sigma_2}\cdot\vec{v})](\frac{1}{\lambda r}+\frac{1}{r^2})e^{-r/\lambda}
	\label{B_6}
\end{equation}
\begin{equation}
	B_{14} = f_{14}\frac{1}{2\pi\gamma}\hat{\sigma_1}\cdot(\hat{\sigma_2}\times\vec{v})(\frac{1}{r})e^{-r/\lambda}
	\label{B_14}
\end{equation}
By integrating the volume of both the spin source and the spin sensor, we calculated the exotic effective magnetic fields sensed by the spin sensor as,
\begin{equation}
	B_{\rm {eff},6} = \frac{1}{V_{\rm\uppercase\expandafter{\romannumeral1}}}  \rho_{\rm pol}\int_{V_{\rm\uppercase\expandafter{\romannumeral1}}} dV\int_{V_{\rm\uppercase\expandafter{\romannumeral2}}}dV' B_6 = \sum_{n=1}^\infty B_{\rm eff,6}^{(n)}\sin(2\pi nf_{\rm vib}t)
	\label{B_eff6}
\end{equation}
\begin{equation}
	B_{\rm {eff},14} = \frac{1}{V_{\rm\uppercase\expandafter{\romannumeral1}}}  \rho_{\rm pol}\int_{V_{\rm\uppercase\expandafter{\romannumeral1}}} dV\int_{V_{\rm\uppercase\expandafter{\romannumeral2}}}dV' B_{14} = \sum_{n=1}^\infty B_{\rm eff,14}^{(n)}\sin(2\pi nf_{\rm vib}t)
	\label{B_eff14}
\end{equation}
where $V_{\rm\uppercase\expandafter{\romannumeral1}}(V_{\rm\uppercase\expandafter{\romannumeral2}})$ stands for the integration volume of the spin sensor (spin source), $\rho_{\rm pol} = 1.14\times10^{23}~$m$^{-3}$ is the experimental magnetic polarized spin density of the spin source.
\begin{figure}[htbp]
	\centering
	\includegraphics[width=0.8\columnwidth]{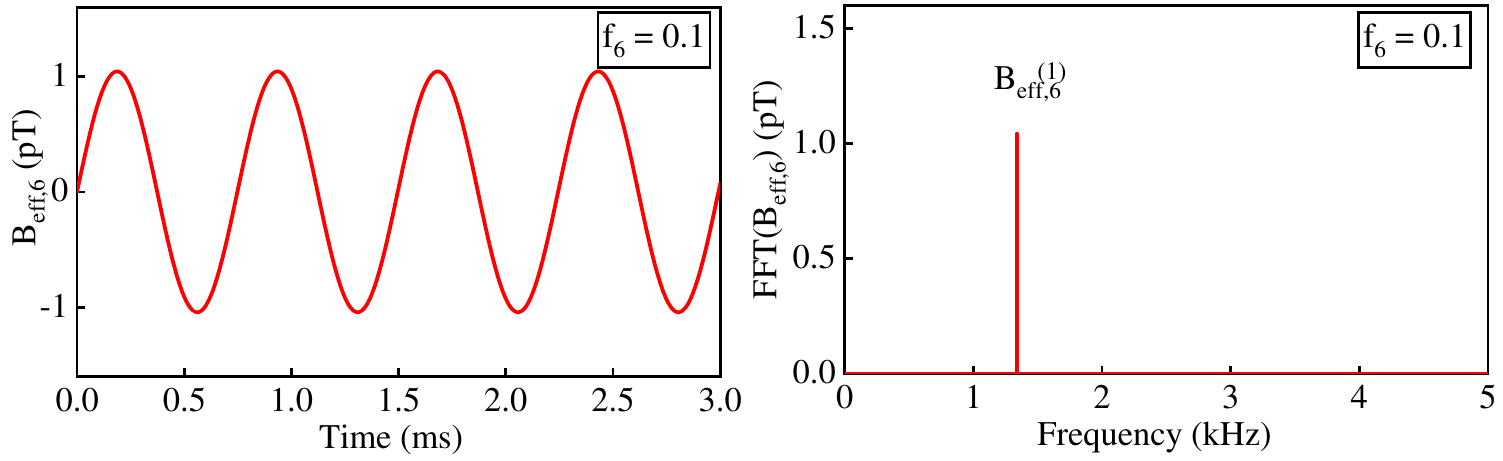}
	\caption{\textbf{Numerical simulation of the exotic effective magnetic field $B_{\rm eff,6}$.} Assume $f_{6}= 0.1$ and $\lambda$ = 1~mm. (a) Time-domain signal of $B_{\rm eff,6}$. (b) Fourier transformation spectrum of $B_{\rm eff,6}$.}
	\label{SMexotic6}
\end{figure}

\begin{figure}[htbp]
	\centering
	\includegraphics[width=0.8\columnwidth]{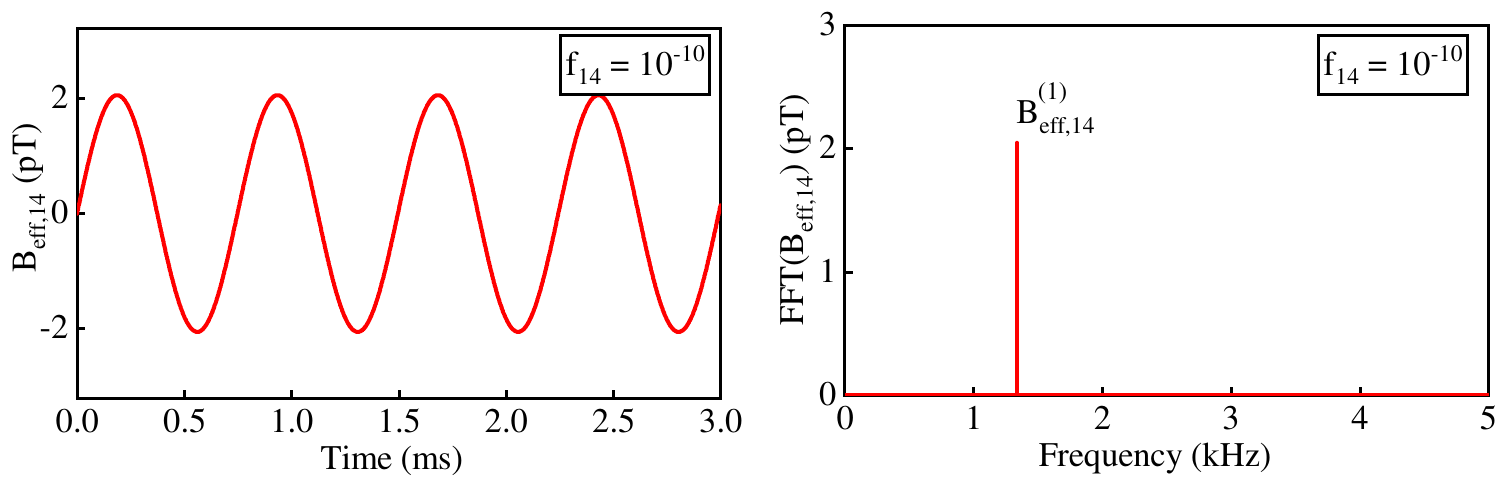}
	\caption{\textbf{Numerical calculation of the exotic effective magnetic field $B_{\rm eff,14}$} Assume $f_{14}=10^{-10}$ and $\lambda$ = 1~mm. (a) Time-domain signal of $B_{\rm eff,14}$. (b) Fourier transformation spectrum of $B_{\rm eff,14}$.}
	\label{SMexotic14}
\end{figure}

The Monte Carlo integral method was utilized to numerically calculate $B_{\rm eff,6}$ and $B_{\rm eff,14}$, with the same algorithm of the calculation of magnetic field due to the magnetic dipole-dipole interaction (see Appendix \ref{Numerical calculations of the magnetic dipole fields}). We take $f_6 =0.1$ and $\lambda$ = 1~mm as an example to present the field stength and the waveform of $B_{\rm eff,6}$ (see left subfigure in Fig. \ref{SMexotic6}). As shown in the right subfigure in Fig. \ref{SMexotic6}, the field strength primarily lies in the first-order harmonic component $B_{\rm {eff},6}^{(1)}$.
We also take $f_{14} = 10^{-10}$ and $\lambda$ = 1~mm as an example to numerical calculate $B_{\rm eff,14}$. The waveform of the exotic effective magnetic field $B_{\rm eff,14}$ is demonstred in the left panel of Fig. \ref{SMexotic14}. As shown in the right panel of Fig. \ref{SMexotic14}, the field strength primarily lies in the first-order harmonic component $B_{\rm {eff},14}^{(1)}$. 

For each distinct force range, the numerical calculations of exotic interactions were performed independently. The procedure is similar to that for the estimation of the magnetic field resulting from the magnetic dipole-dipole interaction, but using the interaction forms described in Eq. \ref{B_6} and Eq. \ref{B_14}. Hence, the geometric factors have been considered in our numerical simulations.

\subsection{Calibration of the phase of the demodulation reference signal}
To search for SSVDIs, the phase of the demodulation reference signal of LIA 2 was carefully calibrated with the same method introduced in Ref.\cite{liang_new_2022,wu_improved_2023}. The calibration procedure was carried out before the experiment searching for exotic signals. A thin copper wire carrying a DC current was stuck to the front section of the piezoelectric bender. The magnetic field $B_{\rm copper}$ generated by the vibrating copper wire was in phase with displacement and was orthogonal to velocity. After the calibration, the phase of the demodulation reference signal of LIA 2 was adjusted to be $\phi = -6.7^{\circ}$ during the experiment searching for SSVDIs. As a result, the displacement-dependent signal corresponds to the in-phase channel, and the velocity-dependent signal $B^{(1)}_{\text{eff}}$ corresponds to the quadrature channel of LIA 2.

\section{Systematic errors}
\label{SM4}

\noindent\textbf{Effects from moving surface charges}

We consider the effects due to the possible moving surface charges of the spin source. First, an AC electric field $E$ will be generated and can couple to the spin sensor through the Stark shift. However, this effect is in phase with the displacement rather than the velocity and can be ignored. Next, we consider the magnetic field generated by the moving charges. The moving spin source with surface charges constitutes a current that generates a magnetic field that is in phase with the velocity. Following the discussion in \cite{PhysRevResearch.4.023162}, in the worst situation where the surface charge density produces an electric field at the threshold of inducing dielectric breakdown in the air, the generated magnetic field is estimated to be $B\approx10$ fT with the parameters in our experiment. This field is much weaker than the statistical error in our experiment, which is about 10 pT. Therefore, the effects of the moving surface charges are expected to be negligible in our experiment.

\noindent\textbf{Effects of the demagnetization factor of the spin source}

We consider the effects of the demagnetization factor determined by the geometry of the spin source. The spin source in our work is a cylinder diamond with a radius of 52 $\upmu$m and a height of 23 $\upmu$m. It is enclosed in Diamond I with a size of $660\times661\times574~\upmu$m$^3$. Hence, the geometry of the diamond will be considered when analyzing the effect of the demagnetization factor. For real magnetic fields, because of the extremely small magnetic susceptibility of the diamond ($\chi_{\text{diamond}}=-2\times{10}^{-5}$\cite{chi_diamond}), the shape effects on the magnetization can be considered insignificant.
To confirm this, we performed numerical simulations of the magnetic dipole field introduced by the spin source on the spin sensor both with and without taking the demagnetization factor into account. In the former case, the magnetic susceptibility of the diamonds was set to $\chi_{\text{diamond}}$. In the latter case, the magnetic susceptibility of the diamonds was set to 0, i.e., the same as the vacuum. The simulated magnetic dipole signals $B_\text{d}^{\left(1\right)}$ obtained by the spin sensor are 6.398 nT in both cases. Therefore, the correction due to the demagnetization factor should be less than 1 pT, which is well below the statistical error of $B_\text{d}^{\left(1\right)}$, and thus can be considered negligible.

\noindent\textbf{Uncertainty in $2R$}

The diameter of the spin source is $2R~=~104(1)~\mu$m, which has been determined according to the fluorescence image with a CCD. The correction to $f_{6}$ is $\pm 0.01$ at $\lambda$ = 1~mm. The correction to $f_{14}$ is $\pm 3\times 10^{-12}$ at $\lambda$ = 1~mm.

\noindent\textbf{Uncertainty in $w$}

The width of the spin sensor is $w = 37.8(7)~\upmu$m, which has been determined according to the fluorescence image. The correction to $f_{6}$ is $\pm 0.001$ at $\lambda$ = 1~mm. The correction to $f_{14}$ is $\pm 1\times 10^{-14}$ at $\lambda$ = 1~mm.

\noindent\textbf{Uncertainty in $h_1$}

The thickness of the layer of the NV spin sensor $h_1$ is estimated by comparing the thickness of the diamond measured before and after the growth of the NV layer. Before the growth of the NV layer, the original thickness of the diamond substrate was 551(1)~$\upmu$m. After the growth of the NV layer, the thickness of the diamond was measured to be 574(1)~$\upmu$m. Thus the thickness of the NV layer of the spin sensor is $h_1=23(1)~\upmu$m. The correction to $f_{6}$ is $\pm0.01$ at $\lambda$ = 1~mm. The correction to $f_{14}$ is $\pm1\times 10^{-12}$ at $\lambda$ = 1~mm.

\noindent\textbf{Uncertainty in $h_2$}

The thickness of the NV layer of the spin source $h_2$ is determined by the same procedure as that of $h_1$. The thickness of the NV layer of the spin source is $h_2=23(1)~\upmu$m. The correction to $f_{6}$ is $\pm0.02$ at $\lambda$ = 1~mm. The correction to $f_{14}$ is $\pm6\times 10^{-12}$ at $\lambda$ = 1~mm.

\noindent\textbf{Uncertainty in $A$}

The vibration amplitude of the spin source is $A=36.7(5)$~nm, measured by a commercial laser vibrometer (Sunny Optical, LV-S01). The correction to $f_{6}$ is $ \pm0.01$ at $\lambda$ = 1~mm. The correction to $f_{14}$ is $ \pm2\times 10^{-12}$ at $\lambda$ = 1~mm.

\noindent\textbf{Uncertainty in $d$}

The distance between the surfaces of the two diamonds is $d=18.5(6)~\upmu$m, which has been measured by an optical microscope. The correction to $f_{6}$ is $ \pm0.01$ at $\lambda$ = 1~mm. The correction to $f_{14}$ is $\pm1\times 10^{-12}$ at $\lambda$ = 1~mm.

\noindent\textbf{Uncertainty in $\phi$}

The uncertainty in $\phi$ was measured to be $4.4^{\circ}$. The correction to $f_{6}$ is from $\pm 0.01$ at $\lambda$ = 1~mm. The correction to $f_{14}$ is from $-2\times 10^{-12}$ to $1\times 10^{-12}$ at $\lambda$ = 1~mm.

\noindent\textbf{Deviation of the spin sensor and the spin source in the x-y plane.}

The deviation of the center of the spin sensor and the center of the spin source was measured to be $46(1)~\upmu$m in $x$-axis direction and $90(2)~\upmu$m in $y$-axis direction according to the CCD images. For deviation in $x$, the correction to $f_{6}$ is from -0.05 to 0.04 at $\lambda$ = 1~mm, and the correction to $f_{14}$ is $\pm1\times 10^{-12}$ at $\lambda$ = 1~mm. For deviation in $y$, the correction to $f_{6}$ is $\pm1\times 10^{-3}$ at $\lambda$ = 1~mm and the correction to $f_{14}$ is $\pm1\times 10^{-13}$ at $\lambda$ = 1~mm.

\noindent\textbf{Uncertainty in Coefficient $\eta$}

The coefficient $\eta$ of the magnetic field to the output of the magnetometer is measured to be $4.1 \pm 0.1  $~V/mT. The correction to $f_{6}$ is $ \pm0.01$ at $\lambda$ = 1~mm. The correction to $f_{14}$ is $\pm3\times 10^{-12}$ at $\lambda$ = 1~mm.

\noindent\textbf{Uncertainty in Polarized density $\rho_{\rm pol}$}

The polarized density $\rho_{\rm pol}$ is obtained to be $(1.14 \pm 0.02)\times10^{23}~$m$^{-3}$. The correction to $f_{6}$ is $ \pm0.01$ at $\lambda$ = 1~mm. The correction to $f_{14}$ is $\pm2\times 10^{-12}$ at $\lambda$ = 1~mm.

\noindent\textbf{Uncertainty in relative angle $\theta$}

The relative angle between the two diamonds $\theta=54.0(3)^\circ$ is measured according to the CCD image. The correction to $f_{6}$ is $ \pm0.01$ at $\lambda$ = 1~mm. The correction to $f_{14}$ is $\pm1\times 10^{-12}$ at $\lambda$ = 1~mm.



%

\end{document}